\documentclass[a4paper, journal]{IEEEtran}
\usepackage{url}
\usepackage[utf8]{inputenc}
\usepackage{xcolor}
\usepackage{amsmath}
\usepackage{amssymb}
\usepackage{algorithm}
\usepackage{algorithmic}

\usepackage[acronyms,nonumberlist,nopostdot,nomain,nogroupskip]{glossaries}
\usepackage{tablefootnote}
\usepackage{booktabs}
\usepackage{tabularx}
\usepackage{tikz}
\usepackage{pgfplots}
\pgfplotsset{compat=newest}
\pgfplotsset{plot coordinates/math parser=false}
\newlength\fheight
\newlength\fwidth
\usetikzlibrary{plotmarks,patterns,decorations.pathreplacing,backgrounds,calc,arrows,arrows.meta,spy,matrix}
\usepgfplotslibrary{patchplots,groupplots}
\usepackage{tikzscale}

\usepackage{multirow}
\usepackage{tkz-kiviat}

\usepackage[font=small]{subcaption}
\usepackage[font=small]{caption}

\usepackage{mathtools}

\usepackage{dblfloatfix}    
\usepackage{colortbl}

\newacronym{3gpp}{3GPP}{3rd Generation Partnership Project}
\newacronym{adc}{ADC}{Analog to Digital Converter}
\newacronym{afbw}{AFBW}{Average Fading Bandwidth}
\newacronym{5g}{5G}{5th generation}
\newacronym{aimd}{AIMD}{Additive Increase Multiplicative Decrease}
\newacronym{am}{AM}{Acknowledged Mode}
\newacronym{amc}{AMC}{Adaptive Modulation and Coding}
\newacronym{aqm}{AQM}{Active Queue Management}
\newacronym{awgn}{AGWN}{Additive White Gaussian Noise}
\newacronym{balia}{BALIA}{Balanced Link Adaptation}
\newacronym{bsr}{BSR}{Buffer Status Report}
\newacronym{bdp}{BDP}{Bandwidth-Delay Product}
\newacronym{bf}{BF}{Beamforming}
\newacronym{isp}{ISP}{Internet Service Provider}
\newacronym{ngmn}{NGMN}{Next Generation Mobile Networks Alliance}
\newacronym{ngnm}{NGNM}{Next Generation Mobile Networks Alliance}
\newacronym{cc}{CC}{Component Carrier}
\newacronym{drb}{DRB}{Data Radio Bearer}
\newacronym{qos}{QoS}{Quality of Service}
\newacronym{ca}{CA}{Carrier Aggregation}
\newacronym{lc}{LC}{Logical Channel}
\newacronym{rnti}{RNTI}{Radio Network Temporary Identifier}
\newacronym{qci}{QCI}{Quality Class Identifier}
\newacronym{cdf}{CDF}{Cumulative Distribution Function}
\newacronym{cn}{CN}{Core Network}
\newacronym{cqi}{CQI}{Channel Quality Information}
\newacronym{cir}{CIR}{Channel Impulse Response}
\newacronym{cp}{CP}{Control Plane}
\newacronym{csirs}{CSI-RS}{Channel State Information - Reference Signal}
\newacronym{dc}{DC}{Dual Connectivity}
\newacronym{dce}{DCE}{Direct Code Execution}
\newacronym{dci}{DCI}{Downlink Control Information}
\newacronym{dl}{DL}{Downlink}
\newacronym{dmr}{DMR}{Deadline Miss Ratio}
\newacronym{dmrs}{DMRS}{DeModulation Reference Signal}
\newacronym{e2e}{E2E}{End-to-End}
\newacronym{ecn}{ECN}{Explicit Congestion Notification}
\newacronym{edf}{EDF}{Earliest Deadline First}
\newacronym{enb}{eNB}{evolved Node Base}
\newacronym{embb}{eMBB}{enhanced Mobile Broadband}
\newacronym{epc}{EPC}{Evolved Packet Core}
\newacronym{es}{ES}{Edge Server}
\newacronym{fdma}{FDMA}{Frequency Division Multiple Access}
\newacronym{fdd}{FDD}{Frequency Division Duplexing}
\newacronym[firstplural=Radio Access Technologies (RATs)]{rat}{RAT}{Radio Access Technology}
\newacronym{fs}{FS}{Fast Switching}
\newacronym{ftp}{FTP}{File Transfer Protocol}
\newacronym{gnb}{gNB}{Next Generation Node Base}
\newacronym{harq}{HARQ}{Hybrid Automatic Repeat reQuest}
\newacronym{hetnet}{HetNet}{Heterogeneous Network}
\newacronym{hh}{HH}{Hard Handover}
\newacronym{hol}{HOL}{Head-of-Line}
\newacronym{ia}{IA}{Initial Access}
\newacronym{imt}{IMT}{International Mobile Telecommunication}
\newacronym{iot}{IoT}{Internet of Things}
\newacronym{lcr}{LCR}{Level Crossing Rate}
\newacronym{lcf}{LCF}{Level Crossing Frequency}
\newacronym{los}{LoS}{Line-of-Sight}
\newacronym{lte}{LTE}{Long Term Evolution}
\newacronym{m2m}{M2M}{Machine to Machine}
\newacronym{mac}{MAC}{Medium Access Control}
\newacronym{mc}{MC}{Multi-Connectivity}
\newacronym{mcs}{MCS}{Modulation and Coding Scheme}
\newacronym{mec}{MEC}{Mobile Edge Cloud}
\newacronym{mi}{MI}{Mutual Information}
\newacronym{mimo}{MIMO}{Multiple Input, Multiple Output}
\newacronym{mmwave}{mmWave}{millimeter wave}
\newacronym{mptcp}{MPTCP}{Multipath TCP}
\newacronym{mr}{MR}{Maximum Rate}
\newacronym{mss}{MSS}{Maximum Segment Size}
\newacronym{mtd}{MTD}{Machine-Type Device}
\newacronym{mtu}{MTU}{Maximum Transmission Unit}
\newacronym{nfv}{NFV}{Network Function Virtualization}
\newacronym{nlos}{NLoS}{Non-Line-of-Sight}
\newacronym{nr}{NR}{New Radio}
\newacronym{o2i}{O2I}{Outdoor-to-Indoor}
\newacronym{ofdm}{OFDM}{Orthogonal Frequency Division Multiplexing}
\newacronym{pdcch}{PDCCH}{Physical Downlonk Control Channel}
\newacronym{pdcp}{PDCP}{Packet Data Convergence Protocol}
\newacronym{pdsch}{PDSCH}{Physical Downlink Shared Channel}
\newacronym{pdu}{PDU}{Packet Data Unit}
\newacronym{pf}{PF}{Proportional Fair}
\newacronym{pgw}{PGW}{Packet Gateway}
\newacronym{phy}{PHY}{Physical}
\newacronym{pbch}{PBCH}{Physical Broadcast Channel}
\newacronym[plural=\gls{mme}s,firstplural=Mobility Management Entities (MMEs)]{mme}{MME}{Mobility Management Entity}
\newacronym{prb}{PRB}{Physical Resource Block}
\newacronym{pss}{PSS}{Primary Synchronization Signal}
\newacronym{pucch}{PUCCH}{Physical Uplink Control Channel}
\newacronym{pusch}{PUSCH}{Physical Uplink Shared Channel}
\newacronym{rach}{RACH}{Random Access Channel}
\newacronym{ran}{RAN}{Radio Access Network}
\newacronym{red}{RED}{Random Early Detection}
\newacronym{rf}{RF}{Radio Frequency}
\newacronym{rlc}{RLC}{Radio Link Control}
\newacronym{rlf}{RLF}{Radio Link Failure}
\newacronym{rrc}{RRC}{Radio Resource Control}
\newacronym{rrm}{RRM}{Radio Resource Management}
\newacronym{rr}{RR}{Round Robin}
\newacronym{rs}{RS}{Remote Server}
\newacronym{rsrp}{RSRP}{Reference Signal Received Power}
\newacronym{rss}{RSS}{Received Signal Strength}
\newacronym{rtt}{RTT}{Round Trip Time}
\newacronym{rw}{RW}{Receive Window}
\newacronym{rx}{RX}{Receiver}
\newacronym{sa}{SA}{standalone}
\newacronym{sack}{SACK}{Selective Acknowledgment}
\newacronym{sap}{SAP}{Service Access Point}
\newacronym{sch}{SCH}{Secondary Cell Handover}
\newacronym{scoot}{SCOOT}{Split Cycle Offset Optimization Technique}
\newacronym{sdma}{SDMA}{Spatial Division Multiple Access}
\newacronym{sinr}{SINR}{Signal-to-Interference-plus-Noise Ratio}
\newacronym{sir}{SIR}{Signal-to-Interference Ratio}
\newacronym{sm}{SM}{Saturation Mode}
\newacronym{snr}{SNR}{Signal-to-Noise Ratio}
\newacronym{son}{SON}{Self-Organizing Network}
\newacronym{ss}{SS}{Synchronization Signal}
\newacronym{srs}{SRS}{Sounding Reference Signal}
\newacronym{sss}{SSS}{Secondary Synchronization Signal}
\newacronym{tb}{TB}{Transport Block}
\newacronym{tcp}{TCP}{Transmission Control Protocol}
\newacronym{tdd}{TDD}{Time Division Duplexing}
\newacronym{tdma}{TDMA}{Time Division Multiple Access}
\newacronym{tfl}{TfL}{Transport for London}
\newacronym{tm}{TM}{Transparent Mode}
\newacronym{trp}{TRP}{Transmitter Receiver Pair}
\newacronym{tti}{TTI}{Transmission Time Interval}
\newacronym{ttt}{TTT}{Time-to-Trigger}
\newacronym{tx}{TX}{Transmitter}
\newacronym{ue}{UE}{User Equipment}
\newacronym{ul}{UL}{Uplink}
\newacronym{uml}{UML}{Unified Modeling Language}
\newacronym{um}{UM}{Unacknowledged Mode}
\newacronym{uma}{UMa}{Urban Macro}
\newacronym{utc}{UTC}{Urban Traffic Control}
\newacronym{vm}{VM}{Virtual Machine}
\newacronym{rsrq}{RSRQ}{Reference Signal Received Quality}
\newacronym{rssi}{RSSI}{Received Signal Strength Indicator}
\newacronym{crs}{CRS}{Cell Reference Signal}
\newacronym{nsa}{NSA}{Non Stand Alone}
\newacronym{mrdc}{MR-DC}{Multi \gls{rat} \gls{dc}}
\newacronym{endc}{EN-DC}{E-UTRAN-\gls{nr} \gls{dc}}
\newacronym{5gc}{5GC}{5G Core}
\newacronym{si}{SI}{Study Item}
\newacronym{iab}{IAB}{Integrated Access and Backhaul}
\newacronym{wf}{WF}{Wired-first}
\newacronym{hqf}{HQF}{Highest-quality-first}
\newacronym{pa}{PA}{Position-aware}
\newacronym{mlr}{MLR}{Maximum-local-rate}
\newacronym{wbf}{WBF}{Wired Bias Function}
\newacronym{mib}{MIB}{Master Information Block}
\newacronym{sib}{SIB}{Secondary Information Block}
\newacronym{kpi}{KPI}{Key Performance Indicator}
\newacronym{ppp}{PPP}{Poisson Point Process}
\newacronym{mpc}{MPC}{Multi Path Component}
\newacronym{rt}{RT}{Ray Tracer}
\newacronym{aoa}{AoA}{Angle of Arrival}
\newacronym{aod}{AoD}{Angle of Departure}
\newacronym{scm}{SCM}{Spatial Channel Model}
\newacronym{inr}{INR}{Interference to Noise Ratio}
\newacronym{qd}{QD}{Quasi Deterministic}
\newacronym{wlan}{WLAN}{Wireless Local Area Network}
\newacronym{cad}{CAD}{Computer-aided Design}
\newacronym{ap}{AP}{Access Point}
\newacronym{sta}{STA}{Station}
\newacronym{urllc}{URLLC}{Ultra-Reliable Low-Latency Communication}
\newacronym{udp}{UDP}{User Datagram Protocol}
\newacronym{laa}{LAA}{Licensed-Assisted Access}

\tikzstyle{startstop} = [rectangle, rounded corners, minimum width=2cm, minimum height=0.5cm,text centered, draw=black]
\tikzstyle{io} = [trapezium, trapezium left angle=70, trapezium right angle=110, minimum width=3cm, minimum height=1cm, text centered, draw=black]
\tikzstyle{process} = [rectangle, minimum width=2cm, minimum height=0.5cm, text centered, draw=black, alignb=center]
\tikzstyle{decision} = [ellipse, minimum width=2cm, minimum height=1cm, text centered, draw=black]
\tikzstyle{arrow} = [thick,<->,>=stealth]
\tikzstyle{line} = [thick,>=stealth]
\tikzstyle{darrow} = [thick,<->,>=stealth,dashed]
\tikzstyle{sarrow} = [thick,->,>=stealth]
\tikzstyle{larrow} = [line width=0.1mm,dashdotted,->,>=stealth]

\makeatletter
\def\grd@save@target#1{%
  \def\grd@target{#1}}
\def\grd@save@start#1{%
  \def\grd@start{#1}}
\tikzset{
  grid with coordinates/.style={
    to path={%
      \pgfextra{%
        \edef\grd@@target{(\tikztotarget)}%
        \tikz@scan@one@point\grd@save@target\grd@@target\relax
        \edef\grd@@start{(\tikztostart)}%
        \tikz@scan@one@point\grd@save@start\grd@@start\relax
        \draw[minor help lines] (\tikztostart) grid (\tikztotarget);
        \draw[major help lines] (\tikztostart) grid (\tikztotarget);
        \grd@start
        \pgfmathsetmacro{\grd@xa}{\the\pgf@x/1cm}
        \pgfmathsetmacro{\grd@ya}{\the\pgf@y/1cm}
        \grd@target
        \pgfmathsetmacro{\grd@xb}{\the\pgf@x/1cm}
        \pgfmathsetmacro{\grd@yb}{\the\pgf@y/1cm}
        \pgfmathsetmacro{\grd@xc}{\grd@xa + \pgfkeysvalueof{/tikz/grid with coordinates/major step x}}
        \pgfmathsetmacro{\grd@yc}{\grd@ya + \pgfkeysvalueof{/tikz/grid with coordinates/major step y}}
        \foreach \x in {\grd@xa,\grd@xc,...,\grd@xb}
        \node[anchor=north] at (\x,\grd@ya) {\pgfmathprintnumber{\x}};
        \foreach \y in {\grd@ya,\grd@yc,...,\grd@yb}
        \node[anchor=east] at (\grd@xa,\y) {\pgfmathprintnumber{\y}};
      }
    }
  },
  minor help lines/.style={
    help lines,
    gray,
    line cap =round,
    xstep=\pgfkeysvalueof{/tikz/grid with coordinates/minor step x},
    ystep=\pgfkeysvalueof{/tikz/grid with coordinates/minor step y}
  },
  major help lines/.style={
    help lines,
    line cap =round,
    line width=\pgfkeysvalueof{/tikz/grid with coordinates/major line width},
    xstep=\pgfkeysvalueof{/tikz/grid with coordinates/major step x},
    ystep=\pgfkeysvalueof{/tikz/grid with coordinates/major step y}
  },
  grid with coordinates/.cd,
  minor step x/.initial=.5,
  minor step y/.initial=.2,
  major step x/.initial=1,
  major step y/.initial=1,
  major line width/.initial=1pt,
}
\makeatother

\makeglossaries
\DeclareUnicodeCharacter{2212}{-} 

\begin{document}

\title{Enabling RAN Slicing Through Carrier Aggregation in \gls{mmwave} Cellular Networks} 

\author{\IEEEauthorblockN{
Matteo Pagin\IEEEauthorrefmark{1},
Francesco Agostini\IEEEauthorrefmark{1},
Tommaso Zugno\IEEEauthorrefmark{1},
Michele Polese\IEEEauthorrefmark{3}, 
Michele Zorzi\IEEEauthorrefmark{1}}\\\vspace{1mm}
  \IEEEauthorblockA{\IEEEauthorrefmark{1}Department of Information Engineering, University of Padova, Italy\\email:\texttt{\{name.surname\}@dei.unipd.it}\\
  \IEEEauthorrefmark{3}Institute for the Wireless Internet of Things, Northeastern University, Boston, MA, USA\\ e-mail: \texttt{m.polese@northeastern.edu}
  }
\thanks{This work was partially supported by the U.S. Department of Commerce/NIST (Award No. 70NANB17H166) and by the CloudVeneto initiative.}
}

\maketitle

\glsunset{nr}
\thispagestyle{empty}

\begin{abstract}
  The ever increasing number of connected devices and of new and heterogeneous mobile use cases implies that 5G cellular systems will face demanding technical challenges. For example, \gls{urllc} and \gls{embb} scenarios present orthogonal \gls{qos} requirements that 5G aims to satisfy with a unified \gls{ran} design. 
  Network slicing and \gls{mmwave} communications have been identified as possible enablers for 5G. They provide, respectively, the necessary scalability and flexibility to adapt the network to each specific use case environment, and low latency and multi-gigabit-per-second wireless links, which tap into a vast, currently unused portion of the spectrum.
  The optimization and integration of these technologies is still an open research challenge, which requires innovations at different layers of the protocol stack. This paper proposes to combine them in a \gls{ran} slicing framework for \glspl{mmwave}, based on carrier aggregation. Notably, we introduce MilliSlice, a cross-carrier scheduling policy that exploits the diversity of the carriers and maximizes their utilization, thus simultaneously guaranteeing high throughput for the \gls{embb} slices and low latency and high reliability for the \gls{urllc} flows.

\end{abstract}

\begin{picture}(0,0)(10,-430)
\put(0,0){
\put(0,0){\footnotesize This paper has been accepted for presentation at IEEE MedComNet 2020. \textcopyright[2020] IEEE.}
\put(0,-10){\footnotesize Please cite it as Matteo Pagin, Francesco Agostini, Tommaso Zugno, Michele Polese, Michele Zorzi, Enabling RAN Slicing Through Carrier Aggregation}
\put(0, -20){\footnotesize in mmWave Cellular Networks, Proc. of the 18th Mediterranean Communication and Computer Networking Conference (MedComNet 2020), Arona, Italy, 2020}}
\end{picture}

\section{Introduction}
Future mobile networks will face numerous technical challenges to jointly satisfy user requirements (i.e., ultra-high throughput, availability and reliability, and low latency) and optimize the \gls{isp} operations (e.g., in terms of cost and energy efficiency)~\cite{rost2017network}. This upsurge of network demands results from (i) the simultaneous increase of mobile terminals; (ii) the diversity of the requested services; and (iii) the rapid evolution of new use cases, such as inter-vehicular communications and smart factory scenarios. Consequently, \gls{5g} networks have been designed to provide connectivity for different classes of services, with orthogonal requirements. For example, a packet error rate of $10^{-4}$ is tolerable in an \gls{embb} system, where the focus is on high throughput; however, when it comes to industrial real time applications, typical target values for reliability are in the order of $10^{-6}$, together with low latency~\cite{frotzscher2014requirements,parvez2018survey}. It follows that the design of new generations of mobile networks should be flexible enough to adapt to the different requirements.

Network slicing, defined by the \gls{ngmn} as the concept of running multiple indipendent logical networks upon a common physical infrastructure, has been proposed as an enabler of flexible 5G networks~\cite{rost2017network}.
Specifically, a network slice is a self-contained, virtualized and independent end-to-end network that allows operators to execute different deployments in parallel, each based on its own architecture~\cite{alliance2016description}. While there have been several research efforts focused on optimizing slicing operations in wired networks (e.g., in the core of cellular networks), and in traditional, sub-6 GHz wireless networks, the state of the art lacks considerations on how this can be applied to the radio access of 5G \gls{mmwave} networks.


\gls{mmwave} communications are another key enabler of 5G, which will exploit the currently unused, vast portion of the radio spectrum that lies in the bands between 30 and 300 GHz to provide multi-gigabit-per-second throughput to mobile users~\cite{Rangan_2014}. Additionally, the small wavelength of \gls{mmwave} signals and the advances in low-power CMOS RF circuits make it possible to install large antenna arrays even in a small form factor, such as that of a smartphone, hence enabling beamforming and \gls{mimo} techniques. However, the harsh propagation characteristics of \gls{mmwave} frequencies and the susceptibility to blockage make reliable, low latency and high throughput communications at such high frequencies very challenging~\cite{rappaport2013millimeter}, and call for the introduction of innovations across all layers of the protocol stack, from the physical and \gls{mac} (e.g., beam management) to the transport and application layers~\cite{zhang2019will}.
In this regard, one promising strategy is multi connectivity~\cite{polese2017jsac,drago2018reliable}, which introduces macro diversity in the \gls{ran}, increasing the robustness with respect to blockage and allowing mobile users to exploit different frequency bands. \gls{ca}, which enables multi-connectivity at the \gls{mac} layer by providing service on multiple links (called \glspl{cc}), is part of the \gls{3gpp} \gls{lte} and \gls{nr} specifications~\cite{pedersen2011carrier,38300} and has been widely deployed to aggregate bandwidth from different portions of the spectrum~\cite{zugno2018integration}.

This paper is one of the first contributions that studies how to effectively combine network slicing and \gls{mmwave} wireless networks, satisfying heterogeneous traffic demands through flexible operations also in these frequency bands. Notably, we focus on how to serve \gls{urllc} and \gls{embb} slices that share the same radio access resources, without compromising the quality of service of the users in either of the two. 
We tackle this problem from an intra-cell perspective, leaving the challenge of guaranteeing seamless service in the presence of user mobility as future work. 
The proposed slicing framework exploits carrier aggregation to (i) distribute the \gls{urllc} and \gls{embb} flows among different carriers, which could effectively act as slices; and (ii) provide frequency diversity, e.g., slices that require high reliability could be allocated in lower portions of the spectrum, which benefit from a reduced pathloss. Additionally, we introduce MilliSlice, a cross-carrier packet scheduling policy that dynamically adapts the dispatching of packets to the different carriers with the goal to maximize the utilization of the resources available in each \gls{cc}, without penalizing the performance requirements of each slice. 

We evaluate the effectiveness of the proposed solution with an open-source, realistic, end-to-end, full-stack network simulator for \glspl{mmwave}~\cite{end2end5gsim} based on ns-3, which features the \gls{3gpp} channel model for mmWave frequencies and a \gls{3gpp}-like protocol stack with carrier aggregation. The results show that, compared to a mmWave network without slicing, the proposed solution reduces the latency of \gls{urllc} flows and increases the throughput of the \gls{embb} streams, hence enhancing the \gls{qos} achieved by both slices at the same time. 



The remainder of this work is organized as follows. In Section \ref{state_art} we provide a brief review of the state of the art regarding network slicing and \gls{ca} solutions. Then, in Section \ref{sol}, we introduce the slicing framework, focusing in particular on its novelty aspects compared to the currently available solutions in the literature. Section \ref{perf_an} provides a simulation-based performance analysis of the presented strategy, and finally we conclude this paper and highlight possible future improvements in Section \ref{conc}.

\section{State of the art}\label{state_art}

This Section will review relevant research efforts for the slicing of the \gls{ran} (Sec.~\ref{sec:slicing}) and carrier aggregation in sub-6~GHz and \gls{mmwave} cellular networks~(Sec.~\ref{sec:ca}).

\subsection{\gls{ran} Slicing}
\label{sec:slicing}

Although introducing network slicing at the \gls{ran} is still challenging, several \gls{5g} initiatives have been pushing for new frameworks to enable network slicing in mobile networks. \cite{ksentini2017toward} proposes a fully programmable network architecture based on a flexible \gls{ran} to enforce network slicing, also implementing a two-level MAC scheduler to share physical resources among slices, obtaining encouraging results in terms of throughput and resource allocation adaptability. Similarly, the authors of~\cite{foukas2017orion} envision fully virtualized \gls{lte} base stations that can be deployed on-the-fly to serve slices with different performance requirements. Moreover, \cite{sallent2017radio} analyzes the \gls{ran} slicing issue in a multi-cell network, presenting four different slicing approaches for splitting the radio resources among slices, and achieving high granularity and flexibility in the assignment of radio resources, as well as satisfactory levels of isolation. Paper~\cite{doro2020slicing} adapts a holistic approach to \gls{ran} slicing, proposing a framework that translates high-level service requests of the operators into a correct mapping of the physical layer resources.
Finally, \cite{garcia2019latency} proposes a novel latency-sensitive 5G \gls{ran} slicing solution for Industry 4.0 scenarios, where stringent latency requirements are common. This proposal, evaluated in industrial scenarios with mixed traffic types, is able to meet the latency requirements of delay-sensitive or time-critical applications, thus improving the \gls{qos} experienced by all traffic types through an efficient allocation of the resources to the slices. However, the schemes that have been proposed so far target traditional sub-6 GHz deployments, while in this work we consider the application of network slicing to \glspl{mmwave} cellular systems.

\subsection{Carrier Aggregation}
\label{sec:ca}
Carrier aggregation is a technique that the \gls{3gpp} has first introduced in the \gls{lte} specifications~\cite{36300}, and extended in NR~\cite{38300}, which enables different \gls{cc}s to operate at different frequencies, and to use different \gls{mcs} or retransmission processes, usually within the same base station.
Moreover, \gls{ca} allows the aggregation of licensed and unlicensed bands with \gls{lte}-U and \gls{laa}~\cite{zhang2015lte}.
The advantages that this approach can bring have been profoundly studied in the literature and eventually even implemented in actual deployments, but mostly within the realm of \gls{lte}-Advanced mobile networks: the employment of \gls{ca} in such scenarios provides an increase of the available per user data-rate (since it can aggregate the radio resources across the spectrum) as well as the  means for an agile interference management~\cite{pedersen2011carrier}.

In 5G cellular systems, the capabilities of \gls{ca} have been extended with the possibility of using up to 16~\cite{38802,38300} carriers with a bandwidth of up to 400~MHz. Moreover, as NR supports \gls{mmwave} communications, it will be possible to combine carriers with different propagation properties (e.g., \gls{mmwave} and sub-6-GHz) or in unlicensed and licensed bands (thanks to the extension of NR-U in the 60 GHz band)~\cite{khan2014carrier,lagen2019new}, in order to increase the throughput and improve the reliability of transmissions~\cite{Rangan_2014}. In our previous work~\cite{zugno2018integration}, we analyzed the performance of different \gls{ca} schemes for \glspl{mmwave} using an end-to-end network simulator~\cite{end2end5gsim}, showing that \gls{ca} improves the throughput of the network, due to the higher resilience to blockage given by macro-diversity and the higher efficiency of a per-carrier scheduling and \gls{mcs} selection.
However, even though the preliminary analysis carried out by means of simulation in \cite{zugno2018integration} shows promising results, the application of this technique to \gls{mmwave}s has not been exhaustively studied so far and presents some open challenges such as the introduction of joint-\gls{cc} schedulers and \gls{mac}-\gls{phy} cross layers approaches.

\section{Efficient mmWave \gls{ran} slicing with \gls{ca}} \label{sol}

In this Section, we will describe the proposed \gls{ran} slicing framework for \gls{mmwave} cellular networks, providing details on how \gls{ca} can be used to perform slicing, and on the cross-carrier scheduling policy that manages to guarantee to each data stream the desired \gls{qos}.

The overall goal is to satisfy the requirements in terms of latency and reliability of \gls{urllc} flows, i.e., over-the-air delay below 1 ms and packet loss smaller than $10^{-6}$, while maximizing the throughput of the \gls{embb} flows that share the same radio interface. We designed the slicing framework to be robust with respect to the number of users per base station, the amount of \gls{embb} traffic, and the configuration of the resource allocation in the access networks.

\begin{figure}[t]
  \centering
  \includegraphics[width=\columnwidth]{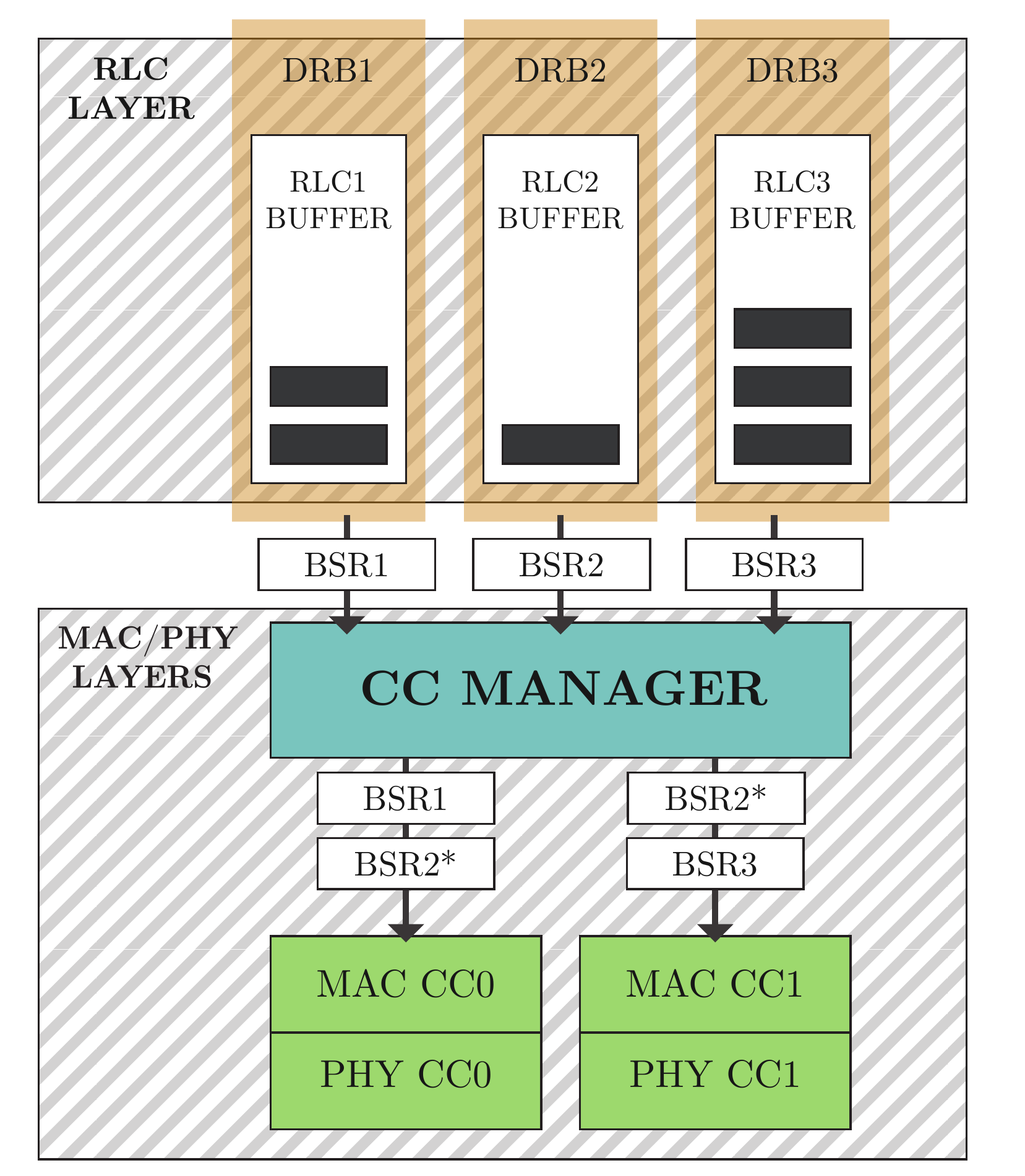}
  \caption{Protocol stack of an \gls{nr} device using \gls{ca}, with a focus on the layers which play a role in the slicing framework. In our proposal the \gls{bsr} messages coming from the above RLC layers are redistributed across the various CCs, possibly depicting different amounts of data compared to their original form (see BSR2 and BSR2*).}
  \label{Fig:log_diag}
\end{figure}

\subsection{\gls{ran} Slicing Through \gls{ca}}

The \gls{ca} technique involves the \gls{phy} and \gls{mac} layers, as well as the interaction between \gls{mac} and \gls{rlc}. Figure~\ref{Fig:log_diag} reports a simplified diagram of the protocol stack with the entities involved in the management of multiple carriers.
During the configuration phase, the base station notifies the availability of one or multiple carrier components, to which the \gls{ue} could connect according to its capabilities.
Once the connection setup is completed, the base station can manage the \glspl{cc} by offloading users to different carriers, or by performing a cross-carrier scheduling for the users connected to multiple \glspl{cc}. As described in the \gls{3gpp} specifications for \gls{nr}~\cite{38300}, the inter-\gls{cc} scheduling in \gls{ca} happens in the highest portion of the \gls{mac} layer, which is interfaced with the different instances of \gls{rlc}.\footnote{In 3GPP networks, each end-to-end data flow is mapped to a \gls{drb}, which, in turn, corresponds to a specific pair of \gls{rlc} and \gls{pdcp} instances.}
The \gls{rlc} periodically sends to the \gls{mac} layer a \gls{bsr}, a report with information about the occupancy of the different buffers (i.e., the size of the transmission and retransmission queues). The \gls{mac} layer then uses the \glspl{bsr} to schedule the radio resources.

In this paper, we propose to adapt the \gls{ca} mechanism to achieve network slicing at the \gls{ran}. As previously highlighted, most of the solutions that have been introduced to perform slicing have been considered for deployment in the core network. Those that have been implemented at the \gls{ran} are based on ad hoc scheduling at the \gls{mac} layer with a single carrier. Unlike these approaches, we propose to implement slicing at \glspl{mmwave} exploiting \gls{ca}, as this solution provides several advantages over the aforementioned single-carrier strategies. 
First of all, it allows the aggregation of multiple carriers, so that the telecom operators could use the available spectrum in a more flexible way. 
Additionally, \gls{ca} enables isolation among the different slices by serving each one with a different carrier.
Finally, it makes it possible to exploit macro diversity, i.e., to allocate flows with different requirements in portions of the spectrum with distinct propagation characteristics. For example, a \gls{cc} with a lower carrier frequency exhibits a smaller pathloss, but, at the same time, may be more constrained in terms of available bandwidth with respect to a \gls{cc} at a higher frequency. This provides a natural fit to serve \gls{urllc} flows in the lower \gls{cc}, as they could benefit from the improved propagation conditions but have limited needs in terms of bandwidth, and the \gls{embb} traffic in the higher portion of the spectrum, trading reliability for a larger bandwidth.
In our work, we follow this principle by always scheduling \gls{urllc} flows in the \gls{cc} with the lowest carrier frequency.

\glsreset{bsr}
\glsreset{drb}

In the proposed slicing framework, when a telecom operator needs to allocate a new \gls{ran} slice for an end-to-end flow with a certain \gls{qos} level, it first checks if the base stations in the area where the slice should be served have \glspl{cc} available to host the slice. If this is the case, it specifies at the \gls{mac} layer of each base station the \gls{qos} requirements corresponding to the specific flow (e.g., whether it is a \gls{urllc} or \gls{embb} flow). These requirements are expressed through a \gls{qci}, i.e., an indicator for the \gls{qos} of each end-to-end flow standardized by the 3GPP~\cite{23501}, associated to the \glspl{bsr} generated at the \gls{rlc} layer.
Eventually, when the slice is operational, the \gls{mac} layer uses the \gls{qci} of the \glspl{bsr} to map it to the proper \gls{cc}. For example, in Figure~\ref{Fig:log_diag}, \gls{rlc}3 serves an \gls{embb} slice, and its \glspl{bsr} are forwarded to \gls{cc}1. Conversely, \gls{rlc}1 is associated to a \gls{urllc} \gls{drb}, and will be scheduled on \gls{cc}0. Notice that in this paper we do not focus on the admission problem, but rather on the optimization of the slice scheduling on the different \glspl{cc}, as we will discuss in the next paragraphs. 

\subsection{Slice-aware Cross-Carrier Scheduling} \label{aware-sched}

As previously mentioned, \gls{ca} enables, in principle, the orthogonal separation of the \gls{urllc} and \gls{embb} slices in different \glspl{cc}. However, as we will highlight in Sec.~\ref{perf_an}, this may lead to inefficiencies in the spectrum utilization, especially in the case where the slices have heterogeneous requirements in terms of bandwidth. In particular, even if the \gls{cc} for the \gls{urllc} slices may be configured with a smaller bandwidth, the datarate difference between \gls{embb} and \gls{urllc} flows, if not properly handled, can lead to the exhaustion of the available capacity in the \gls{embb} slice, with idle resources in the \gls{cc} for \gls{urllc}.

\begin{figure}[t]
  \centering
  \resizebox{0.55\textwidth}{!}{%
    \begin{tikzpicture} 
	  \node[draw=none,fill=none] at (0.05,0.6){\includegraphics[width=.25\linewidth]{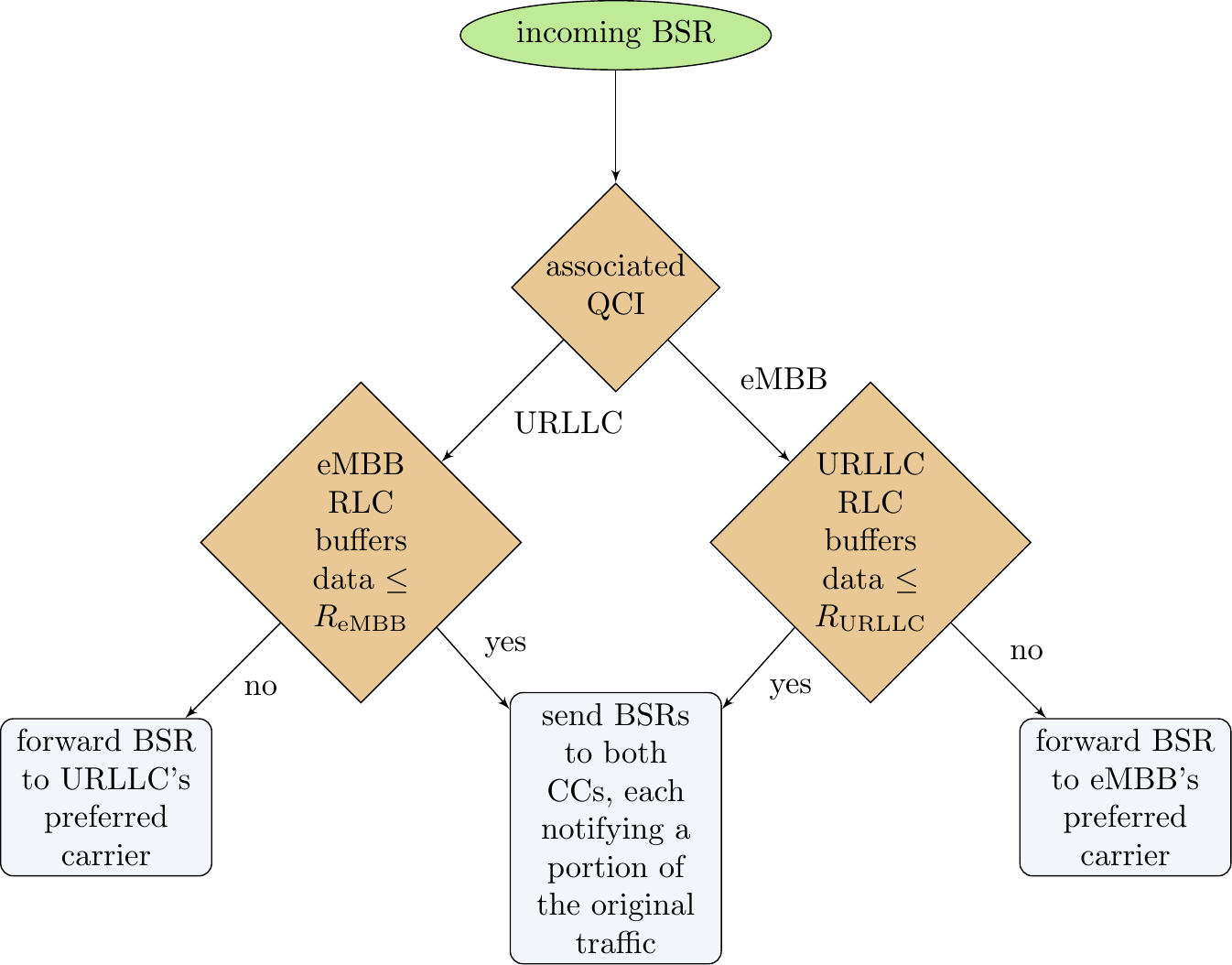}};
	  \end{tikzpicture}
  }%
  \setlength\belowcaptionskip{-.2cm}
  \caption{Flow-chart of the \gls{bsr} scheduling logic}
  \label{Fig:flow_chart}
\end{figure}

Therefore, as part of the proposed \gls{ran} slicing framework, we introduce MilliSlice, a cross-carrier scheduling component whose purpose is to improve the efficiency of the slicing process while avoiding detrimental effects on the \gls{qos} of the \gls{urllc} slices. Referring to Figure~\ref{Fig:log_diag}, this component is deployed in the \gls{cc} manager, thus it does not require any modification in the per-carrier scheduling algorithm that the operator selects for each \gls{cc}.

The slicing framework associates each slice to a \textit{primary} carrier component, following the strategy described in the previous section, and, additionally, to one or more \textit{secondary} \glspl{cc}, in which the slice has a lower priority with respect to the slices that use these \glspl{cc} as primary. The slices that have a low priority on a \gls{cc} will be served in that \gls{cc} only if the flows that use it as primary do not occupy all the available resources. This makes the cross-carrier slicing mechanism adaptive to the load of each slice. Specifically, in the aforementioned case of \gls{urllc} and \gls{embb} sources, the proposed method distributes the data across the \glspl{cc} with the following criteria. The \gls{embb} traffic shall be partially redirected towards its secondary \gls{cc} if and only if the \gls{urllc} buffers (which consider this \gls{cc} as primary) contain less data to transmit than a pre-determined threshold $R_{\rm eMBB}$;
a similar principle applies to the \gls{embb} slices.

The process is based on an adaptive forwarding of the \glspl{bsr} to the different \glspl{cc}, as depicted in the flow chart of Figure~\ref{Fig:flow_chart}. Notably, the component carrier manager, i.e., the entity in charge of splitting the traffic among the different carriers, tracks the buffer occupancy of the \gls{rlc} layers with a sliding window mechanism. Then, once the \gls{rlc} sends a \gls{bsr} to the \gls{mac}, the \gls{cc} manager checks the associated \gls{qci} and, if the buffer occupancy of the secondary carrier is above the predefined threshold, the \gls{bsr} is forwarded to the primary \gls{cc} only. Otherwise, the \gls{bsr} is split across the primary and secondary \glspl{cc}. The pseudocode in Algorithm~\ref{Alg:algo_sched} extends this procedure for a generic number of secondary carrier components.

\begin{algorithm}[t]
  \begin{algorithmic}[1]
    \renewcommand{\algorithmicrequire}{\textbf{Input:}}
    \renewcommand{\algorithmicensure}{\textbf{Output:}}
    \REQUIRE The incoming \glspl{bsr} \texttt{BSR}, the \texttt{Buffer\-Occupancy\-Map} at the \gls{cc} manager, \texttt{QciCcMap}, associating \glspl{qci} to their primary carrier, and the set of thresholds $R$ for each \gls{qci}
    \ENSURE \texttt{ChosenCCs}, a map associating \glspl{cc} and respective \glspl{bsr}
    \STATE Compute the aggregated RLC buffer occupancy (new packets + retransmissions), store it in \texttt{RlcLoads}
    \STATE Consider \texttt{qci}, the \gls{qci} associated with the \gls{bsr} \texttt{BSR} of a specific flow
    \IF {\texttt{Qci} $\in$ \texttt{QciCcMap}}
    \STATE Add the primary CC to the list of available ones
    \STATE \texttt{ChosenCCs}$[$ \texttt{QciCcMap}$[$\texttt{cci}$]$ $]$ $\leftarrow$ \texttt{BSR}
    \STATE Check whether the RLC buffers of the various different slices contain less data than a given threshold, if so add them
    \FORALL{\texttt{entry} $\in$ \texttt{RlcLoads}}
    \STATE \texttt{oQci} $\leftarrow$ \gls{qci} associated with \texttt{entry}
    \IF {\texttt{qci} $\neq$ \texttt{oQci} \textbf{and} \texttt{RlcLoads}$[$ \texttt{oQci} $]$ $ < R_{\rm \texttt{oQci}}$}
    \STATE \texttt{ChosenCCs}$[$ \texttt{QciCcMap}$[$\texttt{oQci}$]$ $]$ $\leftarrow$ \texttt{BSR}
    \ENDIF
    \ENDFOR
    \FORALL {\texttt{cc} $\in$ \texttt{ChosenCCs}}
    \STATE \texttt{ChosenCCs}$[$\texttt{cc}$]$ $\rightarrow$ \texttt{BSR}.\texttt{Tx\-Queue\-Size} $ = $ \texttt{BSR}.\texttt{Tx\-Queue\-Size} $/$ \textbf{size}(\texttt{ChosenCCs})
    \ENDFOR
    \ENDIF
    \RETURN \texttt{ChosenCCs}
  \end{algorithmic}
  \caption{Cross-carrier scheduler implemented in the proposed \gls{ran} slicing framework.}
  \label{Alg:algo_sched}
\end{algorithm}

Furthermore, we choose the carrier operating at lower frequency to be the primary for the \gls{urllc} flow, and set the threshold $R_{\rm eMBB} = 0$, so that the \gls{urllc} traffic is never redistributed across the \glspl{cc} (i.e., it can be served only by its primary \gls{cc}).
This is due to the fact that \gls{urllc} packets would experience a lower average \gls{sinr} on secondary carriers, as the primary is chosen to be the one with the lowest carrier frequency and, additionally, they would be handled with low priority in secondary \glspl{cc}, thus impacting latency and reliability. Conversely, for the \gls{embb} traffic, we set $R_{\rm URLLC} = 1$ packet, so that these slices can be served by the secondary \gls{cc} when the \gls{urllc} RLC buffers are empty.

\section{Performance analysis}\label{perf_an}

This Section will provide insights on the performance that can be achieved using the proposed \gls{ran} slicing framework, after a brief description of the simulator used for the performance evaluation and of the scenario of interest.

\subsection{ns-3 mmWave Module}

The performance analysis has been carried out using simulations with the open-source network simulator ns-3, which allowed us to accurately analyze the end-to-end performance of the proposed slicing framework. Specifically, the simulations are based on the ns-3 mmWave module introduced in~\cite{end2end5gsim}, which features the 3GPP channel model for \glspl{mmwave}~\cite{mmwave3gppchannel}, to stochastically characterize propagation loss, fading, beamforming and interference in the wirless domain, a \gls{3gpp}-like protocol stack for \glspl{gnb} and \glspl{ue}, and, thanks to the integration with ns-3, the possibility of simulating different mobility patterns and the details of the TCP/IP protocol stack.

To implement the slicing framework proposed in this paper, we consider the implementation of \gls{ca} for the ns-3 mmWave module described in~\cite{zugno2018integration}. The \gls{cc} manager that behaves according to the policies described in Sec.~\ref{sec:slicing} is an extension of the \texttt{MmWave\-NoOp\-Component\-Carrier\-Manager} class, which adaptively forwards the \glspl{bsr} from the \gls{rlc} instances to the \glspl{mac} of the various \glspl{cc}. Additionally, we implemented a complete simulation script that can be used to instantiate slicing scenarios and compare different network configurations. The open-source code base associated to this paper is publicly available,\footnote{\url{https://github.com/signetlabdei/millislice}} so that researchers interested in the area of \gls{ran} slicing can use it to further extend this work.

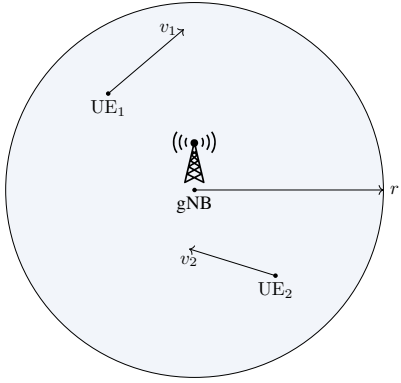
\begin{figure}[t]
  \centering
  \resizebox{0.3\textwidth}{!}{%
    
\definecolor{slate}{RGB}{242, 245, 250}
\begin{tikzpicture}
  \filldraw[fill=slate, draw=black] (0,0) circle (3.5cm);
  \node[draw=none,fill=none] at (0.05,0.6){\includegraphics[width=.25\linewidth]{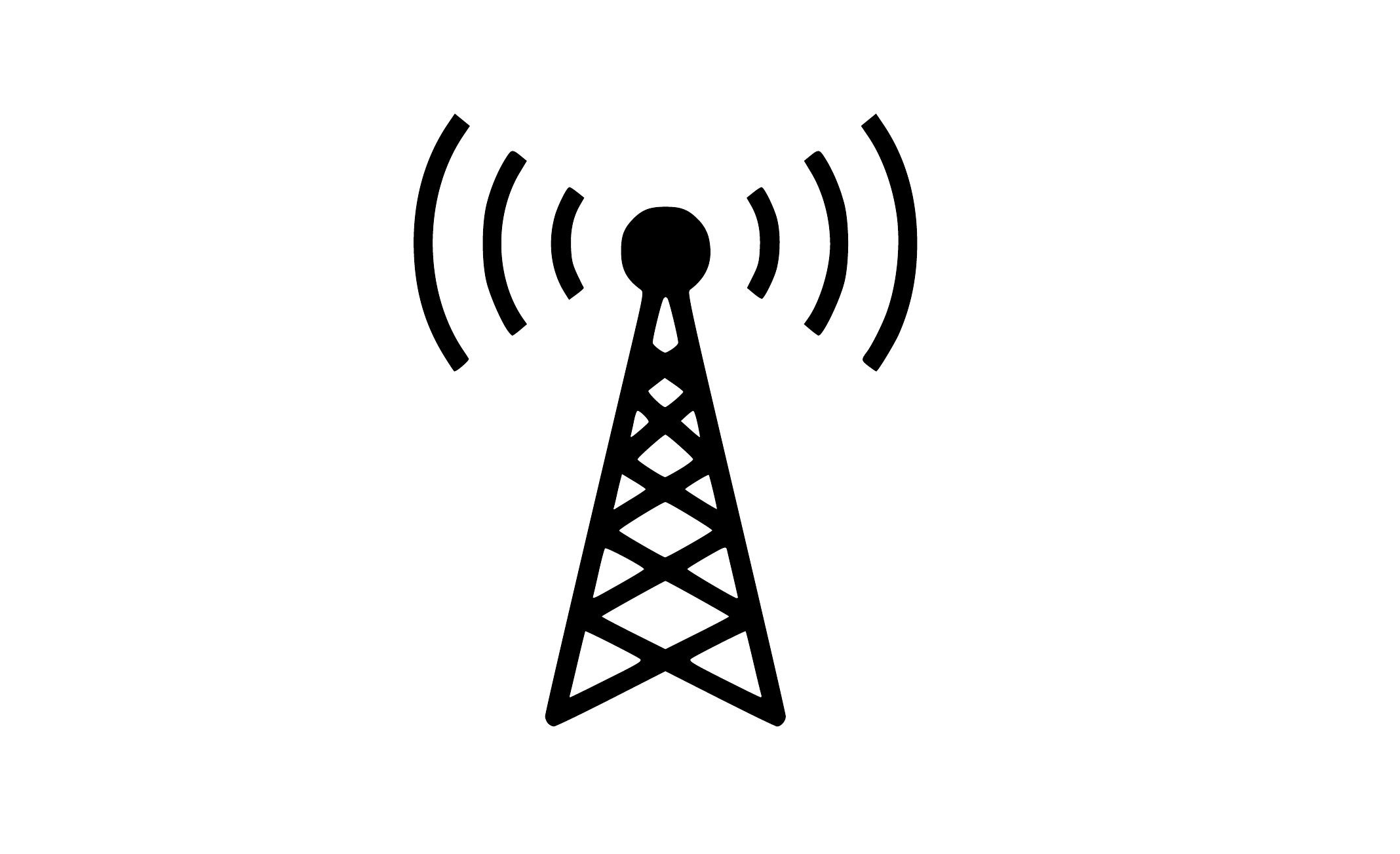}};
  \draw[->] (0,0) -- (3.5,0) node[right] {$r$};
  
  \coordinate (s) at (0,0);
  \fill (s) circle [radius=1.2pt];
  \node [below] at (s) {gNB};

  \coordinate (u1) at (-1.6,1.8);
  \fill (u1) circle [radius=1.2pt];
  \node [below] at (u1) {$\mathrm{UE_{1}}$};
  \draw[->] (u1) -- (-.2,3) node[left] {$v_1$};
  
  \coordinate (u2) at (1.5,-1.6);
  \fill (u2) circle [radius=1.2pt];
  \node [below] at (u2) {$\mathrm{UE_{2}}$};  
  \draw[->] (u2) -- (-.1,-1.1) node[below] {$v_2$};
  
\end{tikzpicture}
  }%
  \caption{Simulation scenario, with \glspl{ue} randomly moving within a circle with radius $r$ around a \gls{gnb}.}
  \label{Fig:sim_scen}
\end{figure}

\subsection{Simulation Scenario and Network Parameters}

\begin{figure*}[t]
  \centering
  \begin{subfigure}[b]{0.3\textwidth}
    \centering
    \setlength\fwidth{0.9\columnwidth}
    \setlength\fheight{0.7\columnwidth}
\begin{tikzpicture}
\pgfplotsset{every tick label/.append style={font=\scriptsize}}

\definecolor{color0}{rgb}{0.12156862745098,0.466666666666667,0.705882352941177}
\definecolor{color1}{rgb}{1,0.498039215686275,0.0549019607843137}
\definecolor{color2}{rgb}{0.172549019607843,0.627450980392157,0.172549019607843}

\begin{axis}[
width=0.951\fwidth,
height=\fheight,
at={(0\fwidth,0\fheight)},
scale only axis,
legend style={at={(0.5,0.95)}, anchor=south, draw=white!80!black, font=\footnotesize},
legend columns=3,
xlabel style={font=\footnotesize},
xlabel={eMBB source rate [Mbit/s]},
xmajorgrids,
xmin=77.265, xmax=163.635,
xtick style={color=white!15!black},
ylabel shift = -3 pt,
ylabel style={font=\footnotesize},
ylabel={URLLC delay [ms]},
ymajorgrids,
ymin=1.48, ymax=1.96250562749417,
ytick style={color=white!15!black},
axis background/.style={fill=white},
]
\path [draw=color0, semithick]
(axis cs:59.8,1.64864615622439)
--(axis cs:59.8,1.85703780462239);

\path [draw=color0, semithick]
(axis cs:80.3,1.67552318275222)
--(axis cs:80.3,1.92611274407972);

\path [draw=color0, semithick]
(axis cs:99.9,1.68012673235906)
--(axis cs:99.9,1.92129479404772);

\path [draw=color0, semithick]
(axis cs:120.5,1.68079368702006)
--(axis cs:120.5,1.91796708088079);

\path [draw=color0, semithick]
(axis cs:138.8,1.67371624983781)
--(axis cs:138.8,1.95235244131296);

\path [draw=color0, semithick]
(axis cs:160.6,1.67716816658394)
--(axis cs:160.6,1.91008740371844);

\path [draw=color1, semithick]
(axis cs:59.8,1.54368284340661)
--(axis cs:59.8,1.55119269531047);

\path [draw=color1, semithick]
(axis cs:80.3,1.54341844967961)
--(axis cs:80.3,1.5508256152588);

\path [draw=color1, semithick]
(axis cs:99.9,1.54351883158696)
--(axis cs:99.9,1.55065200669936);

\path [draw=color1, semithick]
(axis cs:120.5,1.54375241379504)
--(axis cs:120.5,1.55125815663019);

\path [draw=color1, semithick]
(axis cs:138.8,1.5433482080163)
--(axis cs:138.8,1.55115181802071);

\path [draw=color1, semithick]
(axis cs:160.6,1.54341904158332)
--(axis cs:160.6,1.55060585323483);

\path [draw=color2, semithick]
(axis cs:59.8,1.5231655917624)
--(axis cs:59.8,1.53460603849255);

\path [draw=color2, semithick]
(axis cs:80.3,1.56662516785705)
--(axis cs:80.3,1.58522998186971);

\path [draw=color2, semithick]
(axis cs:99.9,1.61195084067224)
--(axis cs:99.9,1.62437140989008);

\path [draw=color2, semithick]
(axis cs:120.5,1.63969092890813)
--(axis cs:120.5,1.64735813556243);

\path [draw=color2, semithick]
(axis cs:138.8,1.64102291576489)
--(axis cs:138.8,1.64897732342334);

\path [draw=color2, semithick]
(axis cs:160.6,1.64138328890668)
--(axis cs:160.6,1.64918353456811);

\addplot [semithick, color0, dash pattern=on 1.5pt off 1pt, mark=square*, mark size=2, mark options={solid,draw=white}]
table {%
80.3 1.77474675841517
99.9 1.77615732123365
120.5 1.77273572647097
138.8 1.77978079820444
160.6 1.77556832429453
};
\addplot [semithick, color1, mark=*, mark size=2, mark options={solid,draw=white}]
table {%
80.3 1.5470326193739
99.9 1.54694268743535
120.5 1.54732866821248
138.8 1.54714705553163
160.6 1.54702756011291
};
\addplot [semithick, color2, dash pattern=on 5pt off 1.5pt, mark=triangle*, mark size=3, mark options={solid,rotate=180,draw=white}]
table {%
80.3 1.57629940495958
99.9 1.6183022105247
120.5 1.64342489831926
138.8 1.64515774123119
160.6 1.64510742179641
};

\end{axis}

\end{tikzpicture}
    \setlength\abovecaptionskip{-.2cm}
    \caption{Average \gls{urllc} delay.}
    \label{Fig:urllc_vs_embb}
  \end{subfigure}
  \hfill
  \begin{subfigure}[b]{0.3\textwidth}
    \centering
    \setlength\fwidth{0.9\columnwidth}
    \setlength\fheight{0.7\columnwidth}
\begin{tikzpicture}
\pgfplotsset{every tick label/.append style={font=\scriptsize}}

\definecolor{color0}{rgb}{0.12156862745098,0.466666666666667,0.705882352941177}
\definecolor{color1}{rgb}{1,0.498039215686275,0.0549019607843137}
\definecolor{color2}{rgb}{0.172549019607843,0.627450980392157,0.172549019607843}

\begin{axis}[
width=0.951\fwidth,
height=\fheight,
at={(0\fwidth,0\fheight)},
scale only axis,
legend style={at={(0.5,1.02)}, anchor=south, draw=white!80!black, font=\footnotesize},
legend columns=3,
xlabel style={font=\footnotesize},
xlabel={eMBB source rate [Mbit/s]},
xmajorgrids,
xmin=77.265, xmax=163.635,
xtick style={color=white!15!black},
ylabel shift = -3 pt,
ylabel style={font=\footnotesize},
ylabel={eMBB throughput [Mbit/s]},
ymajorgrids,
ymin=25, ymax=122.421680544278,
ytick style={color=white!15!black}
]
\path [draw=color0, semithick]
(axis cs:59.8,54.8024843130435)
--(axis cs:59.8,57.1824232626087);

\path [draw=color0, semithick]
(axis cs:80.3,66.2646864139131)
--(axis cs:80.3,70.4274191582609);

\path [draw=color0, semithick]
(axis cs:99.9,75.3266220521739)
--(axis cs:99.9,80.6574616486956);

\path [draw=color0, semithick]
(axis cs:120.5,82.0036189495652)
--(axis cs:120.5,88.507896653913);

\path [draw=color0, semithick]
(axis cs:138.8,84.0303947686957)
--(axis cs:138.8,90.6846110052174);

\path [draw=color0, semithick]
(axis cs:160.6,83.7835675826087)
--(axis cs:160.6,90.7035636869565);

\path [draw=color1, semithick]
(axis cs:59.8,46.285181106087)
--(axis cs:59.8,49.3391595965217);

\path [draw=color1, semithick]
(axis cs:80.3,47.7442085843478)
--(axis cs:80.3,51.1405993182609);

\path [draw=color1, semithick]
(axis cs:99.9,47.4526541913043)
--(axis cs:99.9,50.8956315826087);

\path [draw=color1, semithick]
(axis cs:120.5,47.3495055582609)
--(axis cs:120.5,50.7744089043478);

\path [draw=color1, semithick]
(axis cs:138.8,47.2815538086957)
--(axis cs:138.8,50.5375249808696);

\path [draw=color1, semithick]
(axis cs:160.6,47.0092221217391)
--(axis cs:160.6,50.5135682782609);

\path [draw=color2, semithick]
(axis cs:59.8,59.7632309426087)
--(axis cs:59.8,59.7706580591304);

\path [draw=color2, semithick]
(axis cs:80.3,80.1840766886957)
--(axis cs:80.3,80.2727635478261);

\path [draw=color2, semithick]
(axis cs:99.9,97.8606320417391)
--(axis cs:99.9,99.2896002226087);

\path [draw=color2, semithick]
(axis cs:120.5,102.926477801739)
--(axis cs:120.5,108.264461133913);

\path [draw=color2, semithick]
(axis cs:138.8,102.655557008696)
--(axis cs:138.8,108.207075951304);

\path [draw=color2, semithick]
(axis cs:160.6,102.681992236522)
--(axis cs:160.6,108.039430233043);
\addplot [semithick, color0, dash pattern=on 1pt off 1pt, mark=square*, mark size=2, mark options={solid,draw=white}]
table {%
80.3 68.3525342608696
99.9 78.0235019130435
120.5 85.282459826087
138.8 87.2465541565217
160.6 87.0813428869565
};
\addlegendentry{no CA}
\addplot [semithick, color1, mark=*, mark size=2, mark options={solid,draw=white}]
table {%
80.3 49.3832815304348
99.9 49.2690298434783
120.5 49.0975410086956
138.8 48.9396313043478
160.6 48.7378053565217
};
\addlegendentry{CA, primary only}
\addplot [semithick, color2, dash pattern=on 4pt off 1.5pt, mark=triangle*, mark size=3, mark options={solid,rotate=180,draw=white}]
table {%
80.3 80.2352528695652
99.9 98.6156165565217
120.5 105.604354226087
138.8 105.519246469565
160.6 105.305933913043
};
\addlegendentry{CA, MilliSlice}
\end{axis}

\end{tikzpicture}
    \setlength\abovecaptionskip{-.2cm}
    \caption{Average \gls{embb} throughput.}
    \label{Fig:t_embb_vs_embb}
  \end{subfigure}
  \hfill
  \begin{subfigure}[b]{0.3\textwidth}
    \centering
    \setlength\fwidth{0.9\columnwidth}
    \setlength\fheight{0.7\columnwidth}
\begin{tikzpicture}
\pgfplotsset{every tick label/.append style={font=\scriptsize}}

\definecolor{color0}{rgb}{0.12156862745098,0.466666666666667,0.705882352941177}
\definecolor{color1}{rgb}{1,0.498039215686275,0.0549019607843137}
\definecolor{color2}{rgb}{0.172549019607843,0.627450980392157,0.172549019607843}

\begin{axis}[
width=0.951\fwidth,
height=\fheight,
at={(0\fwidth,0\fheight)},
scale only axis,
legend style={at={(0.5,0.95)}, anchor=south, draw=white!80!black, font=\footnotesize},
legend columns=3,
xlabel style={font=\footnotesize},
xlabel={eMBB source rate [Mbit/s]},
xmajorgrids,
xmin=77.265, xmax=163.635,
xtick style={color=white!15!black},
ylabel shift = -3 pt,
ylabel style={font=\footnotesize},
ylabel={eMBB packet loss},
ymajorgrids,
ymin=-0.0305487049681293, ymax=0.68497104565338,
ytick style={color=white!15!black},
ytick={-0.2,0,0.2,0.4,0.6,0.8},
yticklabels={−0.2,0.0,0.2,0.4,0.6,0.8}
]
\path [draw=color0, semithick]
(axis cs:59.8,0.028475658720786)
--(axis cs:59.8,0.0595522619036086);

\path [draw=color0, semithick]
(axis cs:80.3,0.0862416149556659)
--(axis cs:80.3,0.131106553674808);

\path [draw=color0, semithick]
(axis cs:99.9,0.147832200481039)
--(axis cs:99.9,0.196502701237507);

\path [draw=color0, semithick]
(axis cs:120.5,0.217493940379965)
--(axis cs:120.5,0.26856315410531);

\path [draw=color0, semithick]
(axis cs:138.8,0.287123981590881)
--(axis cs:138.8,0.335437731049382);

\path [draw=color0, semithick]
(axis cs:160.6,0.381805723700596)
--(axis cs:160.6,0.428012841106758);

\path [draw=color1, semithick]
(axis cs:59.8,0.103249182497993)
--(axis cs:59.8,0.144603041268604);

\path [draw=color1, semithick]
(axis cs:80.3,0.255637208854936)
--(axis cs:80.3,0.299682433115374);

\path [draw=color1, semithick]
(axis cs:99.9,0.402616848728556)
--(axis cs:99.9,0.436129557074318);

\path [draw=color1, semithick]
(axis cs:120.5,0.507818191502707)
--(axis cs:120.5,0.535824357594772);

\path [draw=color1, semithick]
(axis cs:138.8,0.573438352090852)
--(axis cs:138.8,0.597421043481854);

\path [draw=color1, semithick]
(axis cs:160.6,0.63277028637245)
--(axis cs:160.6,0.65268235542701);

\path [draw=color2, semithick]
(axis cs:59.8,0)
--(axis cs:59.8,0);

\path [draw=color2, semithick]
(axis cs:80.3,0)
--(axis cs:80.3,0);

\path [draw=color2, semithick]
(axis cs:99.9,0.00241818652855999)
--(axis cs:99.9,0.0103149186711901);

\path [draw=color2, semithick]
(axis cs:120.5,0.0607621893002131)
--(axis cs:120.5,0.0958125018008344);

\path [draw=color2, semithick]
(axis cs:138.8,0.157687413668738)
--(axis cs:138.8,0.196902495550561);

\path [draw=color2, semithick]
(axis cs:160.6,0.275093290552042)
--(axis cs:160.6,0.306980430951873);

\addplot [semithick, color0, dash pattern=on 1pt off 1pt, mark=square*, mark size=2, mark options={solid,draw=white}]
table {%
80.3 0.108352841768525
99.9 0.173471205215441
120.5 0.242231891586444
138.8 0.310378654575593
160.6 0.403854306792449
};
\addplot [semithick, color1, mark=*, mark size=2, mark options={solid,draw=white}]
table {%
80.3 0.277226609189193
99.9 0.420090106631847
120.5 0.520520258806565
138.8 0.585120732167627
160.6 0.642631086066826
};
\addplot [semithick, color2, dash pattern=on 4pt off 1.5pt, mark=triangle*, mark size=3, mark options={solid,rotate=180,draw=white}]
table {%
80.3 0
99.9 0.00581727312726042
120.5 0.0779876363288052
138.8 0.177521169317574
160.6 0.290354301031843
};
\end{axis}

\end{tikzpicture}
    \setlength\abovecaptionskip{-.2cm}
    \caption{Average \gls{embb} packet loss.}
    \label{Fig:l_embb_vs_embb}
  \end{subfigure}
  \caption{Per-user performance metrics achieved for different values of the \gls{embb} source rate; the \gls{urllc} data-rate is fixed at $1.0$ Mbit/s.}
  \label{Fig:embb_vs_embb}
\end{figure*}
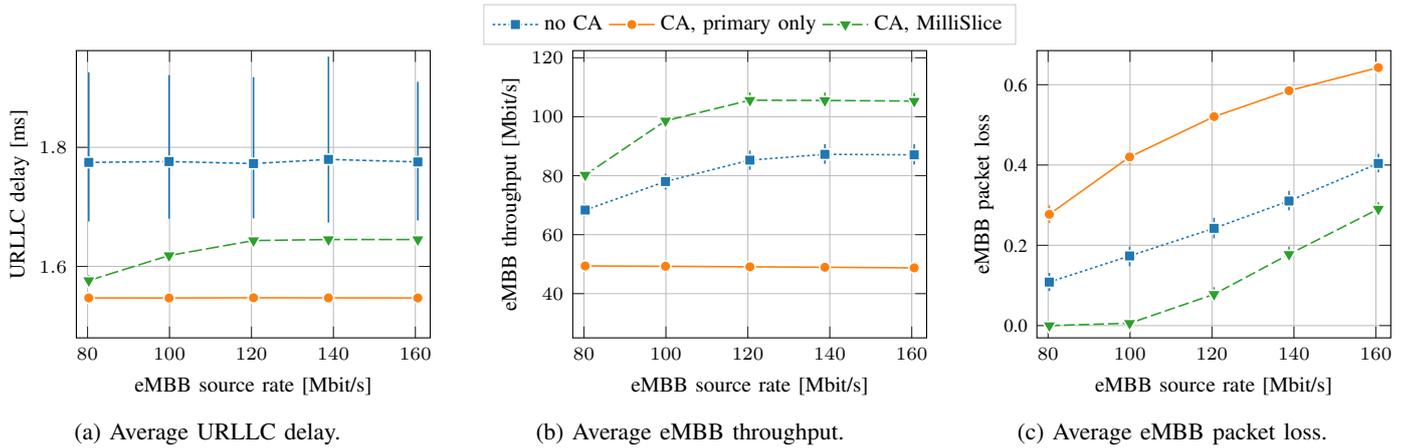

We consider a scenario that models the coverage area of a cell in an urban, densely populated area.
As represented in Figure~\ref{Fig:sim_scen}, the simulation scenario consists of a single cell of radius $200$~m, with one \gls{gnb} at the center and $N_U$ users that are uniformly dropped and move with random speed between $1$ and $10$~m/s.
A remote host connected to the Internet holds \gls{embb} and \gls{urllc} applications, modeled as \gls{udp} sources with different data rates, each generating downlink traffic for a specific user.
The system operates at $28$~GHz with a total bandwidth of $500$~MHz.
In case \gls{ca} is used, an additional carrier component operating at $10$~GHz is added and the overall bandwidth is divided among the two carriers according to the parameter $cc_{ratio}$, which defines the ratio between the bandwidth dedicated to $\mathrm{CC_0}$ and that of $\mathrm{CC_1}$, e.g., when $cc_{ratio}=0.5$, each \gls{cc} is configured with a bandwidth of $250$~MHz.
In our solution, $\mathrm{CC_0}$ will act as the preferred carrier for the \gls{embb} slice, while $\mathrm{CC_1}$ will be dedicated to \gls{urllc} flows.
Finally, as previously mentioned, for this simulation campaign we set $R_{\rm eMBB} = 0$ and $R_{\rm URLLC} =  1$. With this configuration, \gls{urllc} data is never sent to the \gls{embb} \gls{cc}, while \gls{embb} slices can be served by their secondary \gls{cc} only if the \gls{rlc} buffers corresponding to the \gls{urllc} slice are empty. For a more exhaustive list of simulation parameters, please refer to Table \ref{tab:sim_params}.

\begin{table}[t!]
  \renewcommand{\arraystretch}{1.3}
  \caption{Simulation parameters}
  \label{tab:sim_params}
  \centering
  \begin{tabular}{ll}
    \toprule
    Parameter                                 & Value                                \\
    \midrule
    Total System Bandwidth $B$                & $500$ MHz                            \\
    $\mathrm{CC_0}$ center frequency $f_{0}$ & $28$ GHz                             \\
    $\mathrm{CC_1}$ center frequency $f_{1}$ & $10$ GHz                             \\
    \gls{embb} primary \gls{cc}               & $\mathrm{CC_0}$                      \\
    \gls{urllc} primary \gls{cc}              & $\mathrm{CC_1}$                      \\
    RLC Mode                                  & Acknowledged                         \\
    \gls{bsr} timer                           & $1$ ms                               \\
    $cc_{ratio}$                              & 0.5                                  \\
    Number of \gls{urllc} \gls{ue}s           & $10$                                 \\
    Number of \gls{embb} \gls{ue}s            & $10$                                 \\
    \gls{embb} source rate                   & $\left[80, 100, 120, 140, 160\right]$ Mbit/s  \\
    \gls{urllc} source rate                  & $\left[1, 1.5, 2\right]$ Mbit/s      \\
    Radius $r$                                & $200$ m                              \\
    \gls{ue} speed                            & $\mathcal{U} \left[1, 10\right]$ m/s \\
    $R_{\rm URLLC}$                           & 1 packet                             \\
    $R_{\rm eMBB}$                            & 0                                    \\
    \bottomrule
  \end{tabular}
\end{table}

\subsection{Network Configurations and Metrics}
We consider two different baselines to benchmark the performance of the proposed slicing framework. The first (``no CA'' in the plots) is a setup without \gls{ca} and slicing, i.e., with a single carrier with the total system bandwidth $B$. The second (``CA, primary only'' in the plots), instead, is a solution with slicing and \gls{ca}, but without the adaptive cross-carrier scheduling, i.e., in which each slice has only a primary \gls{cc} and cannot use the secondary \gls{cc}.

We evaluated the performance of the proposed framework by analyzing the average end-to-end delay, aggregated throughput and packet loss ratio achieved at the application layer for both the \gls{embb} and \gls{urllc} data flows.
Moreover, to evaluate the per-carrier efficiency in terms of resource utilization, we defined the metric $\eta_{\mathrm{CC_{i}}}$, which represents the portion of the consumed resources with respect of the total available:
\begin{equation}
  \eta_{\mathrm{CC_{i}}} = \frac{tx_{sym}[\mathrm{CC_{i}}]}{t_{sym} \cdot f_{frame} \cdot f_{subframe} \cdot f_{sym}} \times \frac{B_{\mathrm{CC_{i}}}}{B}
\end{equation}
where $tx_{sym}[\mathrm{CC_{i}}]$ is the total number of \gls{ofdm} symbols transmitted through $\mathrm{CC_{i}}$, $t_{sym}$ is the simulation time in seconds, $f_{frame}$ is the number of frames in a second, $f_{subframe}$ is the number of subframes within a frame, and $f_{sym}$ is the number of the \gls{ofdm} symbols which can be transmitted in a subframe.
Moreover, the weight $B_{\mathrm{CC_{i}}}/B$ represents the portion of system bandwidth dedicated to $\mathrm{CC_{i}}$, and is applied to achieve a normalized result.

\subsection{Results}

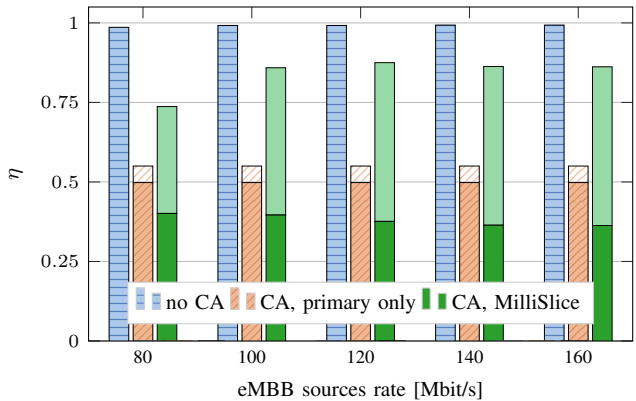
\begin{figure}
  \centering
  \setlength\fwidth{0.85\columnwidth}
  \setlength\fheight{0.5\columnwidth}
  \begin{tikzpicture}

    \pgfplotsset{every tick label/.append style={font=\scriptsize}}

    \definecolor{lightblue}{rgb}{0.672058823529412,0.789705882352941,0.916176470588235}
    \definecolor{lightorange}{rgb}{0.938725490196078,0.718137254901961,0.571078431372549}
    \definecolor{lightgreen}{rgb}{0.596078431372549,0.854901960784314,0.654901960784314}
    \definecolor{darkblue}{rgb}{0.349019607843137,0.490196078431372,0.749019607843137}
    \definecolor{darkorange}{rgb}{0.852941176470588,0.544117647058824,0.370588235294118}
    \definecolor{darkgreen}{rgb}{0.172549019607843,0.627450980392157,0.172549019607843}

    \begin{axis}[ybar stacked,
            bar shift=-9pt,
            bar width=0.18,
            width=0.951\fwidth,
            height=\fheight,
            at={(0\fwidth,0\fheight)},
            scale only axis,
            xlabel style={font=\footnotesize},
            xlabel={eMBB sources rate [Mbit/s]},
            xmin=-0.5, xmax=4.5,
            xtick style={color=white!15!black},
            xtick={0,1,2,3,4},
            xticklabels={80,100,120,140,160},
            ylabel style={font=\footnotesize},
            ylabel={$\eta$},
            ymajorgrids,
            ymin=0, ymax=1.05,
            ytick style={color=white!15!black},
            ytick={0, 0.25, 0.5, 0.75, 1}
        ]
        \addplot [preaction={fill, lightblue}, very thin, pattern={horizontal lines}, pattern color=darkblue] coordinates {
                (0,0.986) (1,0.992) (2,0.992) (3, 0.993) (4, 0.993)
            };
    \end{axis}

    \begin{axis}[
            axis line style={draw=none},
            ybar stacked,
            bar shift=0pt,
            bar width=0.18,
            hide x axis,
            hide y axis,
            width=0.951\fwidth,
            height=\fheight,
            at={(0\fwidth,0\fheight)},
            scale only axis,
            xmajorticks=false,
            ymajorticks=false,
            xmin=-0.5, xmax=4.5,
            ymin=0, ymax=1.05,
        ]
        \addplot [preaction={fill, lightorange}, very thin, pattern={north east lines},
            pattern color=darkorange] coordinates {
                (0,0.498) (1,0.498) (2,0.498) (3, 0.498) (4, 0.498)
            };
        \addplot [pattern={north east lines},
            very thin, pattern color=lightorange] coordinates {
                (0,0.052) (1,0.052) (2,0.052) (3, 0.052) (4, 0.052)
            };
    \end{axis}

    \begin{axis}[
            axis line style={draw=none},
            ybar stacked,
            bar shift=9pt,
            bar width=0.18,
            hide x axis,
            hide y axis,
            width=0.951\fwidth,
            height=\fheight,
            at={(0\fwidth,0\fheight)},
            scale only axis,
            xmajorticks=false,
            xmin=-0.5, xmax=4.5,
            ymin=0, ymax=1.05,
        ]
        \addplot [fill=darkgreen, very thin] coordinates {
                (0,0.401) (1,0.396) (2,0.376) (3,0.364) (4,0.363)
            };
        \addplot [fill=lightgreen, very thin] coordinates {
                (0,0.336) (1,0.463) (2,0.499) (3,0.499) (4,0.499)
            };
    \end{axis}

    \begin{axis}[%
            axis line style={draw=none},
            ybar,
            width=0.951\fwidth,
            height=\fheight,
            at={(0\fwidth,0\fheight)},
            scale only axis,
            xmin=-0.5,
            xmax=2.5,
            ymin=0,
            ymax=1.05,
            xtick=data,
            legend cell align={left},
            legend style={at={(0.5,0.05)}, font=\footnotesize, anchor=south, draw=white!80.0!black},
            axis line style={-},
            hide x axis,
            hide y axis,
            xmajorticks=false,
            ymajorticks=false,
            legend columns=3,
        ]

        \addplot [preaction={fill, lightblue}, pattern={horizontal lines}, pattern color=darkblue]
        coordinates {
                (0,0) (1,0) (2,0)
            };
        \addlegendentry{no CA};

        \addplot [preaction={fill, lightorange}, pattern={north east lines},
            pattern color=darkorange]
        coordinates {
                (0,0) (1,0) (2,0)
            };
        \addlegendentry{CA, primary only};

        \addplot [fill=darkgreen]
        coordinates {
                (0,0) (1,0) (2,0)
            };
        \addlegendentry{CA, MilliSlice};
    \end{axis}

\end{tikzpicture}
  \caption{Evaluation of the resource utilization versus different values of the \gls{embb} source rate and with the \gls{urllc} data-rate fixed at $1.0$ Mbit/s. The darker, bottom portions of the bars represent $\eta_{\mathrm{CC_{0}}}$; the lighter, top ones represent $\eta_{\mathrm{CC_{1}}}$ (when used).}
  \label{Fig:band_vs_embb}
\end{figure}

In Figure~\ref{Fig:embb_vs_embb}, we compare the performance achieved by the three strategies over different values of the \gls{embb} source rate.
Although all the solutions are able to guarantee a reliable delivery of the \gls{urllc} traffic, it can be noticed that the introduction of \gls{ran} slicing by means of carrier aggregation is beneficial for the delay: indeed, both the primary only and MilliSlice solutions show lower \gls{urllc} delay compared with the standard approach. In particular, the lowest delay is reasonably achieved when the two flows are completely isolated, because the usage of dedicated carriers allows \gls{urllc} transmissions to be independently scheduled, without incurring additional delay due to the presence of other \gls{embb} packets in the queue. Moreover, the possibility to employ a carrier operating at a lower frequency ensures a more reliable data delivery,
making it possible to achieve the correct reception of each packet with a smaller number of \gls{mac} and \gls{rlc} layer retransmissions, thus reducing the delay.
However, the advantage that the complete isolation provides for \gls{urllc} traffic comes at the price of sacrificing the \gls{qos} experienced by the \gls{embb} slice, which exhibits lower throughput (Figure~\ref{Fig:t_embb_vs_embb}) and higher packet loss compared with the other solutions (Figure~\ref{Fig:l_embb_vs_embb}).
In this case, the carrier component dedicated to the \gls{embb} flow does not provide enough resources to satisfy the offered traffic and becomes saturated.
Instead, MilliSlice is able to achieve the best performance for the \gls{embb} services while minimizing the \gls{urllc} delay with respect to standard systems, and thus represents a viable solution to achieve network slicing at the \gls{ran} level.
Thanks to an elastic scheduling algorithm, our solution is able to efficiently exploit the available resources by allowing the congested \gls{embb} slice to use the carrier dedicated to the \gls{urllc} flow when idle.
This behavior is confirmed by Figure~\ref{Fig:band_vs_embb}, which represents the resource utilization achieved by the three different approaches, possibly showing the portion used by either CC$_{0}$ (darker) or CC$_{1}$ (lighter) when \gls{ca} is employed.
It can be seen that with MilliSlice more than 80\% of the system resources are exploited and the load is equally distributed among the two carriers.
In contrast, with the primary only approach the carrier dedicated to \gls{urllc} is poorly utilized and about 45\% of the available resources are wasted.
Moreover, the more agile link adaptation provided by \gls{ca} \cite{zugno2018integration} enables MilliSlice to achieve a higher performance gain with respect to the single carrier approach, even using a smaller amount of resources.

\begin{figure*}
  \centering
  \begin{subfigure}[b]{0.3\textwidth}
    \centering
    \setlength\fwidth{0.9\columnwidth}
    \setlength\fheight{0.7\columnwidth}
    \begin{tikzpicture}
\pgfplotsset{every tick label/.append style={font=\scriptsize}}

\definecolor{color0}{rgb}{0.12156862745098,0.466666666666667,0.705882352941177}
\definecolor{color1}{rgb}{1,0.498039215686275,0.0549019607843137}
\definecolor{color2}{rgb}{0.172549019607843,0.627450980392157,0.172549019607843}

\begin{axis}[
width=0.951\fwidth,
height=\fheight,
at={(0\fwidth,0\fheight)},
scale only axis,
legend style={at={(0.5,0.95)}, anchor=south, draw=white!80!black, font=\footnotesize},
legend columns=3,
xlabel style={font=\footnotesize},
xlabel={URLCC source rate [Mbit/s]},
xmajorgrids,
xmin=0.925, xmax=2.075,
xtick style={color=white!15!black},
xtick={0.4,0.6,0.8,1,1.2,1.4,1.6,1.8,2,2.2},
xticklabels={0.4,0.6,0.8,1.0,1.2,1.4,1.6,1.8,2.0,2.2},
ylabel shift = -3 pt,
ylabel style={font=\footnotesize},
ylabel={URLLC delay [ms]},
ymajorgrids,
ymin=1.41522657465033, ymax=2.12853840918966,
ytick style={color=white!15!black},
ytick={1.4,1.6,1.8,2,2.2},
yticklabels={1.4,1.6,1.8,2.0,2.2}
]
\path [draw=color0, semithick]
(axis cs:0.5,1.88091601690832)
--(axis cs:0.5,2.10388442972882);

\path [draw=color0, semithick]
(axis cs:1,1.67783106540372)
--(axis cs:1,1.91890310444452);

\path [draw=color0, semithick]
(axis cs:1.5,1.68174141649266)
--(axis cs:1.5,1.99052157591217);

\path [draw=color0, semithick]
(axis cs:2,1.68176433963217)
--(axis cs:2,2.04461497544463);

\path [draw=color1, semithick]
(axis cs:0.5,1.7639533840417)
--(axis cs:0.5,1.76948038654946);

\path [draw=color1, semithick]
(axis cs:1,1.54326685753039)
--(axis cs:1,1.55081128634453);

\path [draw=color1, semithick]
(axis cs:1.5,1.54339999418907)
--(axis cs:1.5,1.54932563595448);

\path [draw=color1, semithick]
(axis cs:2,1.54538458086844)
--(axis cs:2,1.55468463468789);

\path [draw=color2, semithick]
(axis cs:0.5,1.88290899230978)
--(axis cs:0.5,1.89436436819897);

\path [draw=color2, semithick]
(axis cs:1,1.61165370584431)
--(axis cs:1,1.62445443468118);

\path [draw=color2, semithick]
(axis cs:1.5,1.59420948785988)
--(axis cs:1.5,1.61638256701501);

\path [draw=color2, semithick]
(axis cs:2,1.54189999015226)
--(axis cs:2,1.5914393245299);

\addplot [semithick, color0, dash pattern=on 1pt off 1pt, mark=square*, mark size=2, mark options={solid,draw=white}, forget plot]
table {%
1 1.77615732123365
1.5 1.80258064897827
2 1.82556622989796
};
\addplot [semithick, color1, mark=*, mark size=2, mark options={solid,draw=white}, forget plot]
table {%
1 1.54694268743535
1.5 1.54626115860829
2 1.54986277107481
};
\addplot [semithick, color2, dash pattern=on 4pt off 1.5pt, mark=triangle*, mark size=3, mark options={solid,rotate=180,draw=white}, forget plot]
table {%
1 1.6183022105247
1.5 1.60579805399587
2 1.56743969693939
};

\end{axis}

\end{tikzpicture}
    \setlength\abovecaptionskip{-.2cm}
    \caption{Average \gls{urllc} delay.}
    \label{Fig:urllc_vs_urllc}
  \end{subfigure}
  \hfill
  \begin{subfigure}[b]{0.3\textwidth}
    \centering
    \setlength\fwidth{0.9\columnwidth}
    \setlength\fheight{0.7\columnwidth}
\begin{tikzpicture}
    \pgfplotsset{every tick label/.append style={font=\scriptsize}}

    \definecolor{color0}{rgb}{0.12156862745098,0.466666666666667,0.705882352941177}
    \definecolor{color1}{rgb}{1,0.498039215686275,0.0549019607843137}
    \definecolor{color2}{rgb}{0.172549019607843,0.627450980392157,0.172549019607843}

    \begin{axis}[
    width=0.951\fwidth,
    height=\fheight,
    at={(0\fwidth,0\fheight)},
    scale only axis,
    legend style={at={(0.5,1.02)}, anchor=south, draw=white!80!black, font=\footnotesize},
    legend columns=3,
    xlabel style={font=\footnotesize},
    xlabel={URLCC source rate [Mbit/s]},
    xmajorgrids,
    xmin=0.925, xmax=2.075,
    xtick style={color=white!15!black},
    xtick={0.4,0.6,0.8,1,1.2,1.4,1.6,1.8,2,2.2},
    xticklabels={0.4,0.6,0.8,1.0,1.2,1.4,1.6,1.8,2.0,2.2},
    ylabel shift = -3 pt,
    ylabel style={font=\footnotesize},
    ylabel={eMBB throughput [Mbit/s]},
    ymajorgrids,
    ymin=25, ymax=127.646334280348,
    ytick style={color=white!15!black}
    ]

\path [draw=color0, semithick]
(axis cs:0.5,74.8610486539131)
--(axis cs:0.5,80.4749585808696);

\path [draw=color0, semithick]
(axis cs:1,75.2813058226087)
--(axis cs:1,80.4351189704348);

\path [draw=color0, semithick]
(axis cs:1.5,74.39871488)
--(axis cs:1.5,79.8377845982609);

\path [draw=color0, semithick]
(axis cs:2,73.7813802295652)
--(axis cs:2,79.2352434086957);

\path [draw=color1, semithick]
(axis cs:0.5,47.3434238886956)
--(axis cs:0.5,50.6668272417391);

\path [draw=color1, semithick]
(axis cs:1,47.5275157147826)
--(axis cs:1,50.9584165843478);

\path [draw=color1, semithick]
(axis cs:1.5,47.1768193113043)
--(axis cs:1.5,50.5797667617391);

\path [draw=color1, semithick]
(axis cs:2,46.9199469078261)
--(axis cs:2,50.1130798747826);

\path [draw=color2, semithick]
(axis cs:0.5,96.0681100243478)
--(axis cs:0.5,98.1311007165217);

\path [draw=color2, semithick]
(axis cs:1,97.8814386086956)
--(axis cs:1,99.273218226087);

\path [draw=color2, semithick]
(axis cs:1.5,98.5038213565217)
--(axis cs:1.5,99.6459050295652);

\path [draw=color2, semithick]
(axis cs:2,71.9249543791304)
--(axis cs:2,92.0986557217391);

\addplot [semithick, color0, dash pattern=on 1pt off 1pt, mark=square*, mark size=2, mark options={solid,draw=white}]
table {%
1 78.0235019130435
1.5 77.2595712
2 76.5127279304348
};
\addlegendentry{no CA}
\addplot [semithick, color1, mark=*, mark size=2, mark options={solid,draw=white}]
table {%
1 49.2690298434783
1.5 48.8329572173913
2 48.5729502608696
};
\addlegendentry{CA, primary only}
\addplot [semithick, color2, dash pattern=on 4pt off 1.5pt, mark=triangle*, mark size=3, mark options={solid,rotate=180,draw=white}]
table {%
1 98.6156165565217
1.5 99.1707314086957
2 82.2676613565217
};
\addlegendentry{CA, MilliSlice}
\end{axis}

\end{tikzpicture}
    \setlength\abovecaptionskip{-.2cm}
    \caption{Average \gls{embb} throughput.}
    \label{Fig:t_embb_vs_urllc}
  \end{subfigure}
  \hfill
  \begin{subfigure}[b]{0.3\textwidth}
    \centering
    \setlength\fwidth{0.9\columnwidth}
    \setlength\fheight{0.7\columnwidth}
\begin{tikzpicture}
\pgfplotsset{every tick label/.append style={font=\scriptsize}}
    
\definecolor{color0}{rgb}{0.12156862745098,0.466666666666667,0.705882352941177}
\definecolor{color1}{rgb}{1,0.498039215686275,0.0549019607843137}
\definecolor{color2}{rgb}{0.172549019607843,0.627450980392157,0.172549019607843}
    
\begin{axis}[
width=0.951\fwidth,
height=\fheight,
at={(0\fwidth,0\fheight)},
scale only axis,
legend style={at={(0.5,0.95)}, anchor=south, draw=white!80!black, font=\footnotesize},
legend columns=3,
xlabel style={font=\footnotesize},
xlabel={URLCC source rate [Mbit/s]},
xmajorgrids,
xmin=0.925, xmax=2.075,
xtick style={color=white!15!black},
xtick={0.4,0.6,0.8,1,1.2,1.4,1.6,1.8,2,2.2},
xticklabels={0.4,0.6,0.8,1.0,1.2,1.4,1.6,1.8,2.0,2.2},
ylabel shift = -3 pt,
ylabel style={font=\footnotesize},
ylabel={eMBB packet loss },
ymajorgrids,
ymin=-0.0805154093996748, ymax=0.625473190211857,
ytick style={color=white!15!black},
ytick={0.0, 0.1,0.2,0.3,0.4,0.5,0.6,0.7},
yticklabels={0.0, 0.1,0.2,0.3,0.4,0.5,0.6,0.7}
]
\path [draw=color0, semithick]
(axis cs:0.5,0.15311092899752)
--(axis cs:0.5,0.200949198463079);

\path [draw=color0, semithick]
(axis cs:1,0.152433018801289)
--(axis cs:1,0.200002189050526);

\path [draw=color0, semithick]
(axis cs:1.5,0.156179709625334)
--(axis cs:1.5,0.20511986626179);

\path [draw=color0, semithick]
(axis cs:2,0.162153301210117)
--(axis cs:2,0.209431186698618);

\path [draw=color1, semithick]
(axis cs:0.5,0.406833126745548)
--(axis cs:0.5,0.441038238273152);

\path [draw=color1, semithick]
(axis cs:1,0.403312599889564)
--(axis cs:1,0.436337253595612);

\path [draw=color1, semithick]
(axis cs:1.5,0.40677278332118)
--(axis cs:1.5,0.441894241184445);

\path [draw=color1, semithick]
(axis cs:2,0.411679438402175)
--(axis cs:2,0.443768268635668);

\path [draw=color2, semithick]
(axis cs:0.5,0.00868041538362303)
--(axis cs:0.5,0.0225725572245067);

\path [draw=color2, semithick]
(axis cs:1,0.00225366536333789)
--(axis cs:1,0.0104056746191871);

\path [draw=color2, semithick]
(axis cs:1.5,0.000138686860043924)
--(axis cs:1.5,0.00649774160116112);

\path [draw=color2, semithick]
(axis cs:2,0.0617713562754719)
--(axis cs:2,0.228295757314587);

\addplot [semithick, color0, dash pattern=on 1pt off 1pt, mark=square*, mark size=2, mark options={solid,draw=white}, forget plot]
table {%
1 0.173471205215441
1.5 0.180134070002684
2 0.186179795890355
};
\addplot [semithick, color1, mark=*, mark size=2, mark options={solid,draw=white}, forget plot]
table {%
1 0.420090106631847
1.5 0.424454018662887
2 0.427060369632872
};
\addplot [semithick, color2, dash pattern=on 4pt off 1.5pt, mark=triangle*, mark size=3, mark options={solid,rotate=180,draw=white}, forget plot]
table {%
1 0.00581727312726042
1.5 0.00244395735739932
2 0.14387429575704
};

\end{axis}

\end{tikzpicture}
    \setlength\abovecaptionskip{-.2cm}
    \caption{Avergage \gls{embb} packet loss ratio.}
    \label{Fig:l_embb_vs_urllc}
  \end{subfigure}
  \caption{Per-user performance metrics achieved for different values of the \gls{urllc} source rate; the \gls{embb} data-rate is fixed at $100$ Mbit/s.}
  \label{Fig:embb_vs_urllc}
\end{figure*}
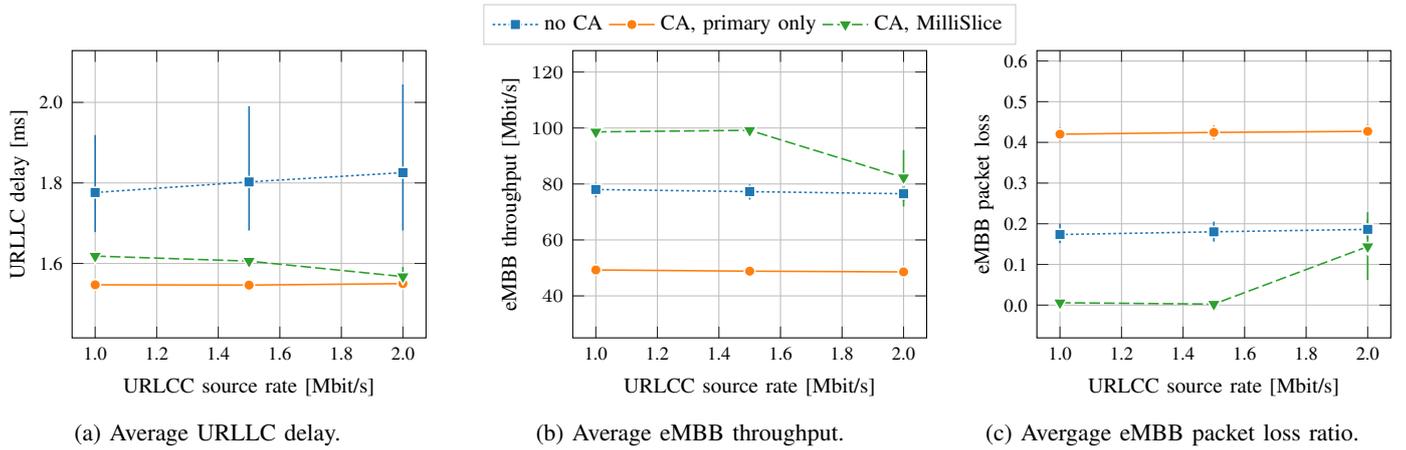

If, on the other hand, we analyze the effectiveness of MilliSlice across different \gls{urllc} source data-rates, we can recognize a similar general trend of the various metrics: in Figure \ref{Fig:embb_vs_urllc} our solution exhibits a higher throughput and lower packet loss for the \gls{embb} flow compared with the other solutions, coupled with a reduction of the \gls{urllc} delay with respect to the single carrier approach.

However, by observing Figures~\ref{Fig:t_embb_vs_urllc} and \ref{Fig:l_embb_vs_urllc}, it can be noticed that the gain introduced by MilliSlice decreases when increasing the rate of the \gls{urllc} sources.
This phenomenon can be interpreted as follows: as the amount of \gls{urllc} traffic increases, the \gls{bsr} arrival occurrences indicating that the respective \gls{rlc} buffers are empty significantly decrease; in turn, given our threshold choices, this results in a reduction of the scheduling instances implementing a redistribution of the traffic across the different \gls{cc}s, hence the inability to sustain the \gls{embb} demands. Nevertheless, we deem it possible to significantly enhance the effectiveness of our \gls{cc} usage policy by coupling such strategy with an ad hoc, slicing-oriented, \gls{mac} layer scheduling, as such choice would enable different and specifically more aggressive \gls{bsr} redistribution strategies by the component carrier manager.

\begin{figure}
  \centering
    \setlength\fwidth{0.9\columnwidth}
    \setlength\fheight{0.5\columnwidth}
\begin{tikzpicture}
\pgfplotsset{every tick label/.append style={font=\scriptsize}}

\definecolor{lightblue}{rgb}{0.578,0.785,0.902}
\definecolor{blue}{rgb}{0.336,0.57,0.82}
\definecolor{darkblue}{rgb}{0.0,0.117,0.308}

\begin{axis}[
 width=0.951\fwidth,
height=\fheight,
at={(0\fwidth,0\fheight)},
scale only axis,
legend style={at={(0.5,1.02)}, anchor=south, draw=white!80!black, font=\footnotesize},
legend columns=3,
xlabel style={font=\footnotesize},
xlabel={Aggregated eMBB throughput [Mbit/s]},
xmajorgrids,
xmin=406.38877529334, xmax=1093.047126195,
xtick style={color=white!15!black},
ylabel style={font=\footnotesize},
ylabel={URLLC delay [ms]},
ymajorgrids,
ymin=1, ymax=5.4,
ytick style={color=white!15!black}
]
\addplot[
  scatter,
  only marks,
  scatter src=explicit,
  scatter/classes={1={lightblue}, 2={blue}, 3={darkblue}},
  mark=otimes,
  mark size=5,
  forget plot,
]
table[x=x,y=y, meta=class]{%
x                      y              class
499.201958956522 1.51263726062566 3
437.607646608696 1.5262757758262 2
473.501250782609 1.85394929419404 1
};

\addplot[
  scatter,
  only marks,
  scatter src=explicit,
  scatter/classes={1={lightblue}, 2={blue}, 3={darkblue}},
  forget plot,
  mark=10-pointed star,
  mark size=5,
]
table[x=x,y=y, meta=class]{%
x                      y              class
987.749598608696 1.61797132538314 3
483.491127652174 1.54812933842854 2
756.091592347826 1.8982780347999 1
};

\addplot[
  scatter,
  only marks,
  scatter src=explicit,
  scatter/classes={1={lightblue}, 2={blue}, 3={darkblue}},
  mark size=5,
]
table[x=x,y=y, meta=class]{%
x                      y              class
1061.81792278261 1.68903426930748 3
455.376628869565 1.5635130355957 2
801.274123130435 4.87477177912112 1
};
\legend{no CA,CA primary only,CA MilliSlice}


\end{axis}

\begin{axis}[
 width=0.951\fwidth,
height=\fheight,
at={(0\fwidth,0\fheight)},
scale only axis,
legend style={at={(0.98,0.98)}, anchor=north east, draw=white!80!black, font=\footnotesize},
xlabel style={font=\footnotesize},
xmajorgrids,
xmin=406.38877529334, xmax=1093.047126195,
xtick style={color=white!15!black},
ylabel style={font=\footnotesize},
ymajorgrids,
ymin=1, ymax=5.4,
ytick style={color=white!15!black},
hide y axis,
hide x axis,
]

\addplot[
  scatter,
  only marks,
  scatter src=explicit,
  mark size=5,
  color=black,
]
table[x=x,y=y]{%
x                      y
-20 -20 
};
\addlegendentry{15 eMBB users}

\addplot[
  scatter,
  only marks,
  scatter src=explicit,
  mark size=5,
  mark=10-pointed star,
  color=black,
]
table[x=x,y=y]{%
x                      y
-20 -21 
};
\addlegendentry{10 eMBB users}

\addplot[
  scatter,
  only marks,
  scatter src=explicit,
  mark size=5,
  mark=otimes,
  color=black,
]
table[x=x,y=y]{%
x                      y
-20 -22
};
\addlegendentry{5 eMBB users}

\end{axis}

\end{tikzpicture}
    \setlength\belowcaptionskip{-.3cm}
    \caption{Average \gls{urllc} delay versus aggregated \gls{embb} throughput.}
    \label{Fig:t_d_versus_numEmbbUes}
\end{figure}
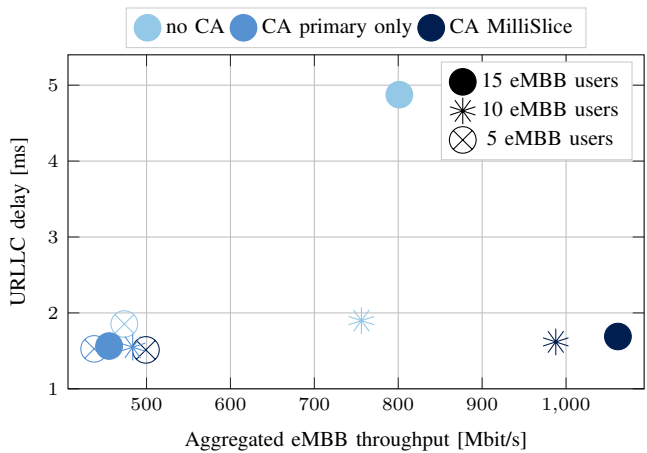
To evaluate the robustness of the proposed scheduling algorithm to possible scenario variations, we analyzed the system behavior by varying the number of users. The results are shown in Figure~\ref{Fig:t_d_versus_numEmbbUes}, in which each point represents the achieved performance in terms of average \gls{urllc} delay and aggregated \gls{embb} throughput when considering a certain number of users.
One one hand, in the single carrier case, the lack of any slicing strategy makes the \gls{urllc} performance susceptible to the increase of the number of \gls{embb} sources. On the other hand, the static carrier assignment isolates the two traffic types onto their favored carrier and lacks any degree of adaptability to the offered \gls{embb} traffic. Intead, MilliSlice manages to scale well and sustain different numbers of \gls{embb} sources while keeping the \gls{urllc} delay under $2$~ms.



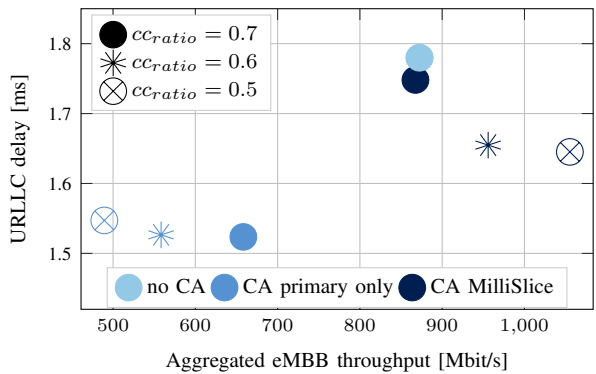
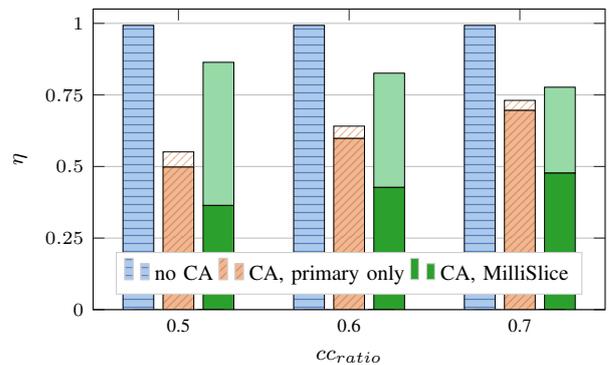
\begin{figure*}
  \centering
  \begin{subfigure}[t]{\columnwidth}
    \centering
    \setlength\fwidth{0.8\columnwidth}
    \setlength\fheight{0.45\columnwidth}
\begin{tikzpicture}
\pgfplotsset{every tick label/.append style={font=\scriptsize}}

\definecolor{lightblue}{rgb}{0.578,0.785,0.902}
\definecolor{blue}{rgb}{0.336,0.57,0.82}
\definecolor{darkblue}{rgb}{0.0,0.117,0.308}

\begin{axis}[
width=0.951\fwidth,
height=\fheight,
at={(0\fwidth,0\fheight)},
scale only axis,
legend style={at={(0.5,0.14)}, anchor=north, draw=white!80!black, font=\footnotesize},
legend columns=3,
xlabel style={font=\footnotesize},
xlabel={Aggregated eMBB throughput [Mbit/s]},
xmajorgrids,
xmin=461.098664559062, xmax=1083.49011318007,
xtick style={color=white!15!black},
ylabel style={font=\footnotesize},
ylabel={URLLC delay [ms]},
ymajorgrids,
ymin=1.42, ymax=1.85,
ytick style={color=white!15!black},
ytick={1.4,1.5,1.6,1.7,1.8,1.9},
yticklabels={1.4,1.5,1.6,1.7,1.8,1.9}
]

\addplot[
  scatter,
  only marks,
  scatter src=explicit,
  scatter/classes={1={lightblue}, 2={blue}, 3={darkblue}},
  mark=otimes,
  mark size=5,
  forget plot,
]
table[x=x,y=y, meta=class]{%
x                      y              class
1055.19246469565 1.64515774123119 3
489.396313043478 1.54714705553163 2
872.465541565217 1.77978079820444 1
};

\addplot[
  scatter,
  only marks,
  scatter src=explicit,
  scatter/classes={1={lightblue}, 2={blue}, 3={darkblue}},
  forget plot,
  mark=10-pointed star,
  mark size=5,
]
table[x=x,y=y, meta=class]{%
x                      y              class
955.868204521739 1.65489482752011 3
558.393344 1.52648218342001 2
872.465541565217 1.77978079820444 1
};

\addplot[
  scatter,
  only marks,
  scatter src=explicit,
  scatter/classes={1={lightblue}, 2={blue}, 3={darkblue}},
  mark size=5,
]
table[x=x,y=y, meta=class]{%
x                      y              class
867.496648347826 1.74804166045435 3
658.371539478261 1.52356648982872 2
872.465541565217 1.77978079820444 1
};
\legend{no CA,CA primary only,CA MilliSlice}

\end{axis}

\begin{axis}[
width=0.951\fwidth,
height=\fheight,
at={(0\fwidth,0\fheight)},
scale only axis,
legend style={at={(0.02,0.98)}, anchor=north west, draw=white!80!black, font=\footnotesize},
xlabel style={font=\footnotesize},
xmajorgrids,
xmin=461.098664559062, xmax=1083.49011318007,
xtick style={color=white!15!black},
ylabel style={font=\footnotesize},
ymajorgrids,
ymin=1.42, ymax=1.85,
ytick style={color=white!15!black},
ytick={1.4,1.5,1.6,1.7,1.8,1.9},
yticklabels={1.4,1.5,1.6,1.7,1.8,1.9},
hide x axis,
hide y axis,
]

\addplot[
  scatter,
  only marks,
  scatter src=explicit,
  mark size=5,
  color=black,
]
table[x=x,y=y]{%
x                      y
-20 -20 
};
\addlegendentry{$cc_{ratio} = 0.7$}

\addplot[
  scatter,
  only marks,
  scatter src=explicit,
  mark size=5,
  mark=10-pointed star,
  color=black,
]
table[x=x,y=y]{%
x                      y
-20 -21 
};
\addlegendentry{$cc_{ratio} = 0.6$}

\addplot[
  scatter,
  only marks,
  scatter src=explicit,
  mark size=5,
  mark=otimes,
  color=black,
]
table[x=x,y=y]{%
x                      y
-20 -22
};
\addlegendentry{$cc_{ratio} = 0.5$}

\end{axis}

\end{tikzpicture}
    \caption{Average \gls{urllc} delay versus aggregated \gls{embb} throughput.}
    \label{Fig:u_vs_d_vs_ccratio}
  \end{subfigure}
  \hfill
  \begin{subfigure}[t]{\columnwidth}
    \centering
    \setlength\fwidth{0.8\columnwidth}
    \setlength\fheight{0.45\columnwidth}
    \begin{tikzpicture}

    \pgfplotsset{every tick label/.append style={font=\scriptsize}}

    \definecolor{lightblue}{rgb}{0.672058823529412,0.789705882352941,0.916176470588235}
    \definecolor{lightorange}{rgb}{0.938725490196078,0.718137254901961,0.571078431372549}
    \definecolor{lightgreen}{rgb}{0.596078431372549,0.854901960784314,0.654901960784314}
    \definecolor{darkblue}{rgb}{0.349019607843137,0.490196078431372,0.749019607843137}
    \definecolor{darkorange}{rgb}{0.852941176470588,0.544117647058824,0.370588235294118}
    \definecolor{darkgreen}{rgb}{0.172549019607843,0.627450980392157,0.172549019607843}

    \begin{axis}[ybar stacked,
            bar shift=-15pt,
            bar width=0.18,
            width=0.951\fwidth,
            height=\fheight,
            at={(0\fwidth,0\fheight)},
            scale only axis,
            xlabel style={font=\footnotesize},
            xlabel={$cc_{ratio}$},
            xmin=-0.5, xmax=2.5,
            xtick style={color=white!15!black},
            xtick={0,1,2},
            xticklabels={0.5,0.6,0.7},
            ylabel style={font=\footnotesize},
            ylabel={$\eta$},
            ymajorgrids,
            ymin=0, ymax=1.05,
            ytick style={color=white!15!black},
            ytick={0, 0.25, 0.5, 0.75, 1}
        ]
        \addplot [preaction={fill, lightblue}, very thin, pattern={horizontal lines}, pattern color=darkblue] coordinates {
                (0,0.993) (1,0.993) (2,0.993)
            };
    \end{axis}

    \begin{axis}[
            axis line style={draw=none},
            ybar stacked,
            bar shift=0pt,
            bar width=0.18,
            hide x axis,
            hide y axis,
            width=0.951\fwidth,
            height=\fheight,
            at={(0\fwidth,0\fheight)},
            scale only axis,
            xmajorticks=false,
            ymajorticks=false,
            xmin=-0.5, xmax=2.5,
            ymin=0, ymax=1.05,
        ]
        \addplot [preaction={fill, lightorange}, very thin, pattern={north east lines},
            pattern color=darkorange] coordinates {
                (0,0.498) (1,0.598) (2,0.696)
            };
        \addplot [pattern={north east lines},
            very thin, pattern color=lightorange] coordinates {
                (0,0.053) (1,0.043) (2,0.035)
            };
    \end{axis}

    \begin{axis}[
            axis line style={draw=none},
            ybar stacked,
            bar shift=15pt,
            bar width=0.18,
            hide x axis,
            hide y axis,
            width=0.951\fwidth,
            height=\fheight,
            at={(0\fwidth,0\fheight)},
            scale only axis,
            xmajorticks=false,
            xmin=-0.5, xmax=2.5,
            ymin=0, ymax=1.05,
        ]
        \addplot [fill=darkgreen, very thin] coordinates {
                (0,0.364) (1,0.427) (2,0.477)
            };
        \addplot [fill=lightgreen, very thin] coordinates {
                (0,0.5) (1,0.399) (2,0.3)
            };
    \end{axis}

    \begin{axis}[%
            axis line style={draw=none},
            ybar,
            width=0.951\fwidth,
            height=\fheight,
            at={(0\fwidth,0\fheight)},
            scale only axis,
            xmin=-0.5,
            xmax=2.5,
            ymin=0,
            ymax=1.05,
            xtick=data,
            legend cell align={left},
            legend style={at={(0.5,0.05)}, font=\footnotesize, anchor=south, draw=white!80.0!black},
            axis line style={-},
            hide x axis,
            hide y axis,
            xmajorticks=false,
            ymajorticks=false,
            legend columns=3,
        ]

        \addplot [preaction={fill, lightblue}, pattern={horizontal lines}, pattern color=darkblue]
        coordinates {
                (0,0) (1,0) (2,0)
            };
        \addlegendentry{no CA};

        \addplot [preaction={fill, lightorange}, pattern={north east lines},
            pattern color=darkorange]
        coordinates {
                (0,0) (1,0) (2,0)
            };
        \addlegendentry{CA, primary only};

        \addplot [fill=darkgreen]
        coordinates {
                (0,0) (1,0) (2,0)
            };
        \addlegendentry{CA, MilliSlice};
    \end{axis}

\end{tikzpicture}
    \setlength\belowcaptionskip{.1cm}
    \caption{Evaluation of the resource utilization over different values of $cc_{ratio}$. The darker, bottom portions of the bars represent $\\eta_{\mathrm{CC_{0}}}$; the lighter, top ones represent $\eta_{\mathrm{CC_{1}}}$ (when used).}
    \label{Fig:band_vs_ccratio}
  \end{subfigure}
  \caption{Evaluation of the system behavior when changing distribution of the system bandwidth among the carrier components by means of the parameter $cc_{ratio}$. The \gls{urllc} sources rate is fixed to $1.0$ Mbit/s, while the \gls{embb} sources rate is fixed to $140$ Mbit/s.}
  \label{Fig:perf_vs_ccratio}
\end{figure*}

Finally, in Figure~\ref{Fig:u_vs_d_vs_ccratio} we can observe how our proposed solution shows poor adaptation capabilities with respect to a variation of the $cc_{ratio}$: as one of the carriers starts to gain possession of most of the bandwidth, the simplicity of our traffic redistribution strategy, coupled with the lack of ad hoc \gls{mac} layer scheduling solutions, starts to show some limitations, even though it still outperforms the other solutions. In particular, such loss in the effectiveness of our policy is driven by a sub-optimal exploitation of the system bandwidth: as depicted by Figure~\ref{Fig:band_vs_ccratio}, the \gls{cc} whose dedicated resources are lower tends to be backlogged, while the other one does not absorb as much traffic as it would be capable of.

\section{Conclusions and Future Work}\label{conc}
The variety of services that 5G networks will have to support requires both the exploitation of previously unexplored portions of the spectrum (i.e., the \gls{mmwave} frequencies) and of additional flexibility in the \gls{ran} configuration. In this paper, we proposed to combine two enablers of 5G networks, i.e., network slicing and carrier aggregation, to support in the same radio interface simultaneous transmission of \gls{urllc} and \gls{embb} traffic flows. Specifically, we proposed a simple but effective policy for the distribution of the various traffic flows among different slices, mapped across multiple carrier components, also exploiting the diversity of the different frequency bands available at mmWaves. We implemented such solution in the ns-3 mmWave module, and carried out an extensive simulation campaign, benchmarking our solution with a number of metrics against two different baseline policies. The promising results and the effectiveness of the proposed solution showed that network slicing through carrier aggregation, especially when coupled with an adaptive cross-carrier scheduling, can sustain heterogeneous 5G requirements.

Future work will focus on more refined solutions, aimed at improving the operations of the schedulers that operate at the carrier component level, to make them aware of the kind of traffic flow they need to support, and to integrate more advanced policies in the proposed slicing framework. 
Moreover, we will analyze more in detail the performance of our solution when scaling the number of flows, designing smart admission policies to efficiently exploit the available resources while ensuring the desired \gls{qos}.  

\bibliographystyle{IEEEtran}
\bibliography{../bibl.bib}

\end{document}